\documentclass[a4paper,12pt,onecolumn,draftclsnofoot]{IEEEtran}
\usepackage{amssymb}
\usepackage{tabularx}
\usepackage{graphicx}
\usepackage{caption}
\usepackage{subcaption}
\usepackage[tbtags]{amsmath}
\usepackage{amsfonts}
\usepackage{mathrsfs}
\usepackage{multirow}
\usepackage[english]{babel}
\usepackage{cite}
\usepackage{url}
\usepackage{hyperref}
\hypersetup{
  colorlinks=true,
  linkcolor=red,
  linktoc=all,
  citecolor=blue,
  bookmarksnumbered=true
}
\usepackage{cleveref}
\usepackage{units}
\usepackage[section]{placeins}

\usepackage[colorinlistoftodos]{todonotes}

\IEEEoverridecommandlockouts

\begin{document}
\title{COVID-19 Epidemic Study II:\\
Phased Emergence From the Lockdown in Mumbai}
\author{
\IEEEauthorblockN{
IISc-TIFR Covid-19 City-Scale Simulation Team}
\thanks{
Authors in alphabetical order of last names:
Prahladh Harsha\IEEEauthorrefmark{2},
Sandeep Juneja\IEEEauthorrefmark{2},
Preetam Patil\IEEEauthorrefmark{1},
Nihesh Rathod\IEEEauthorrefmark{1},
Ramprasad Saptharishi\IEEEauthorrefmark{2},
A. Y. Sarath\IEEEauthorrefmark{1},
Sharad Sriram\IEEEauthorrefmark{1},
Piyush Srivastava\IEEEauthorrefmark{2},
Rajesh Sundaresan\IEEEauthorrefmark{1},
Nidhin Koshy Vaidhiyan\IEEEauthorrefmark{1}}
\thanks{{\bf Corresponding author}: Sandeep Juneja, juneja@tifr.res.in}\\
\thanks{PP, NR, AYS, SS, NKV from IISc were supported by the IISc-Cisco Centre for Networked Intelligence, Indian Institute of Science. RSun was supported by the IISc-Cisco Centre for Networked Intelligence, the Robert Bosch Centre for Cyber-Physical Systems, and the Department of Electrical Communication Engineering, Indian Institute of Science.}
\thanks{TIFR authors acknowledge support of the Department of Atomic Energy, Government of India, under project no. 12-R\&D-TFR-5.01-0500.}
\IEEEauthorblockA{
\IEEEauthorrefmark{1}Indian Institute of Science, Bengaluru\\
\IEEEauthorrefmark{2}TIFR, Mumbai}\\
{04 June 2020}
}
\maketitle

\vspace*{-2.5cm}
\section{Summary}

The nation-wide lockdown starting 25 March 2020, aimed at suppressing the spread of the COVID-19 disease, was extended until 31 May 2020 in three subsequent orders by the Government of India. The extended lockdown has had significant social and economic consequences and `lockdown fatigue' has likely set in. Phased reopening began from 01 June 2020 onwards; however, containment zones will continue to be in lockdown\footnote{See the Ministry of Home Affairs Order dated 30 May 2020 on re-opening \cite{20200530-MHAOrder}}.
Mumbai, one of the most crowded cities in the world, has witnessed both the largest number of cases and deaths among all the cities in India (41986 positive cases and 1368 deaths as of 02 June 2020~\cite{20200602-BMCtwitter}). Many tough decisions are going to be made on re-opening in the next few days.

In ~\cite{report1}, we presented an agent-based city-scale simulator\footnote{Source code available at \url{https://github.com/cni-iisc/epidemic-simulator}.} (ABCS) to model the progression and spread of the
infection in large metropolises  like Mumbai and Bengaluru.  As discussed in \cite{report1}, ABCS is a useful tool to model interactions of city residents at an individual level and to capture the impact of non-pharmaceutical interventions on the infection spread.

In this report we focus on Mumbai. Using our simulator, we consider some plausible scenarios for phased emergence of Mumbai from the lockdown, 01 June 2020 onwards. These include phased and gradual opening of the industry, partial opening of public transportation (modelling of infection spread in suburban trains), impact of containment zones on controlling infections, and the role of compliance with respect to various intervention measures including use of masks, case isolation, home quarantine, etc.

The main takeaway of our simulation results is that a phased opening of workplaces, say at a conservative attendance level of 20 to 33\%, is a good way to restart economic activity while ensuring that the city's medical care capacity remains adequate to handle the possible rise in the number of COVID-19 patients in June and July. In arriving at this level of activity, we assumed that the city has 8,000 beds for COVID-19 patients with severe symptoms requiring hospitalization in addition to 1,500 ICU beds for COVID-19 patients.

The above suggestions are based on indirect estimates of the impact of suburban trains, since these trains have not been in operation since 23 March 2020. A key benefit of restricting workplaces to start operation at a low level is that it keeps the occupancy on the suburban trains and buses at a similar low level, thereby reducing infection spread in crowded public transit systems. (See \cite{harris2020subways} for some speculation on the impact of the New York City (NYC) subway system on COVID-19 spread in NYC.) Furthermore, by starting slowly, the city gets a chance to assess the impact of infection spread in trains and respond appropriately in terms of future relaxations or restrictions on train occupancy.

Our specific recommendation for trains is that occupancy be restricted to about 20\% for the first few weeks with enforcement of strict physical distancing and compulsory wearing of face covers. Further relaxations could be based on the observed disease trends. Standard operating procedures for long-distance trains such as monitoring the temperature of all commuters at entry points of suburban trains and regular disinfection routines should be followed.


We re-emphasize that our simulator is intended primarily as a tool for comparing the effectiveness of different non-medical interventions to assist decision making. In particular, the simulator is not intended as a tool for predicting absolute numerical values of COVID-19 cases. We also recognize that many of the non-pharmaceutical interventions considered in our study, especially when they remain implemented over a long duration, may lead to important social and economic concerns and consequences, beyond their effect on the evolution of the epidemic.  The modelling of such effects still remains beyond the scope of our simulator.

Similar to our previous report, we emphasize that this report has been prepared to help researchers and public health officials understand the effectiveness of social distancing interventions related to COVID-19. The report should not be used for medical diagnostic, prognostic or treatment purposes or for guidance on personal travel plans.

\pagebreak[2]

\section{An Exploration of Specific Unlocking Strategies}

India's national lockdown, which began on 25 March 2020 and was originally scheduled to end on 14 April 2020, was extended in three subsequent orders to 31 May 2020 (Lockdown phase 2: 14 April -- 3 May 2020, Lockdown phase 3: 3 May -- 18 May 2020 and Lockdown phase 4: 18 May -- 31 May 2020). To mitigate hardship to the public during this extended lockdown period, the Ministry of Home Affairs, in its orders dated 15 April 2020 \cite{20200415-MHAOrder}, 1 May 2020 \cite{20200501-MHAOrder}, and 17 May 2020 \cite{20200517-MHAOrder}, allowed certain activities to take place subject to operationalization by states, union territories and district administrations. These local administrations, if they deem appropriate, may impose stricter measures by not allowing these activities to take place. Phased reopening began from 01 June 2020 onwards; however, containment zones will continue to be in lockdown\footnote{See the Ministry of Home Affairs Order dated 30 May 2020 on re-opening \cite{20200530-MHAOrder}.}.
Mumbai has witnessed both the largest number of cases and deaths among all the cities in India (41986 positive cases and 1368 deaths as of 02 June 2020~\cite{20200602-BMCtwitter}). Consequently, Mumbai is likely to be operating with containment zone restrictions, and many tough decisions will be made on re-opening in the next few days.

In this work, we use our agent-based simulator, described in detail in \cite{report1}, to model these relaxations, restrictions and interventions and perform comparative analyses across these measures. We use the synthetic model of the city of Mumbai developed in the first part of the work as a test-bed for our study.

Greater Mumbai (consisting of Mumbai and Suburban Mumbai) has a population of about 1.24 crores (12.4 million) and a population density of roughly 21,000 per km$^2$ \cite{mumbai_census2011_A,suburban_mumbai_census2011_A} making it one of the densest cities in the world. Further, about 53\%~\cite{MCGM_census2011_report}   of Mumbai lives in cramped dwellings with shared sanitation facilities where the population density may be 5 to 10 times larger than other parts of the city.  In addition, crowded suburban trains are the lifeline of the city where the suburban railway system serves more than 78 lakh (7.8 million) passenger trips daily, in normal times. It is generally believed that the infection spreads faster in denser areas, due to increased contacts in these areas. Given these factors, the public health threat in Mumbai is particularly acute. The importance of modelling the effect of infection spread arising from the gradual opening and relaxation of lockdown measures, for a city like Mumbai, cannot then be over-emphasized.

\vspace*{.1in}

{\noindent \bf Agent-based city simulator (ABCS):} As described in detail in \cite{report1}, our agent-based simulator creates a synthetic model of about  1.24 crore (12.4 million) residents of Mumbai that matches the city population ward-wise, and matches the numbers employed, numbers in schools, commute distances, etc. This is done by suitably populating households, schools, and workplaces with people. Several interaction spaces including households, local communities, schools, workplaces, trains, etc. are then modelled to realistically capture the spread of infection. The synthetic city is then seeded with infections to match the observed fatalities till April 10. The infective individuals expose the susceptible individuals to the disease through their interactions in the various interaction spaces. The disease then incrementally evolves in time. The tool helps keep track of the number infected in the city as well as the disease progression within an infected individual. A person infected by the disease may remain asymptomatic and recover, or may develop symptoms. A symptomatic person may recover or may develop severe symptoms and be hospitalized. A patient hospitalized may recover or may become critical. A critical patient may recover or may become deceased. The disease progression parameters are based on \cite{verity2020estimates}.

\vspace{0.1in}

{\noindent \bf Strengths and weaknesses of ABCS:} The advantage of the agent-based city simulator is first and foremost the ability to take into account individuals' features (age, medical conditions, socio-economic factors, etc.) and study their impact on disease epidemiology. A second and perhaps equally important advantage is the capability to model the impact of complex non-medical interventions on the evolution of the disease, e.g., full or partial lockdowns at the required geospatial granularity, closures of school and workplaces, restrictions in containment zones, etc., and inform decision making during the course of the epidemic. These interventions reduce the opportunities for infection-spreading contacts between the infective individuals and the susceptible individuals in the city. Our simulator has been programmed for various time-varying interventions in the synthetically created city. We can then compare the effectiveness of these interventions in terms of their resulting hospitalizations and fatalities.

The accuracy of our model naturally relies on the goodness of our various modelling assumptions and the validity of the underlying parameters. As discussed in \cite{report1}, the key underlying parameters are set to match the deaths from infections that arose in the no-intervention pre-lockdown scenario.

One drawback of ABCS is the many other parameters that need to be tuned. Also, the disease progression statistics that we use are not based upon data observed from India (since a scientific study that considers Indian data is surprisingly not yet available). This increases the model uncertainty and questions the numerical values produced by our model. Perhaps, it is reasonable to assume that the modelling uncertainty is similar across different interventions, and hence we believe our model is much better suited at comparing different interventions in terms of their future trends in epidemic evolution.

\vspace{0.1in}

{\noindent \bf Scenarios considered:} In this work, we perform a variety of simulations of these allowed relaxations and restrictions to enable local administrators to evaluate the public health impact of these measures. To this end, we include features such as adaptive modelling of containment zones, trains, case isolation, home quarantine, etc.

We begin with a brief summary of these new features added to our simulator.

\begin{enumerate}
  \item {\bf Phased opening of industries:} While several restrictions remained in force until 18 May 2020 (e.g., domestic and international air travel\footnote{Limited domestic air travel has been resumed since 25 May 2020.}, passenger movement by trains\footnote{A limited number of long-distance trains have begun to allow for people working in the informal sector to travel and return to their homes.}, buses for public transport, suburban and metro rail services, inter-district and inter-state movement of individuals, closure of educational, training and coaching institutions, closure of industrial and commercial activities, hospitality services, entertainment, shopping, religious places of worship, and restrictions on religious congregations) relaxations were allowable for other activities. In particular, all essential services were operational while Government services have been running at 5\% since 18 May 2020 in Mumbai. We provide a comparative study of the public health impact across the following different scenarios of the phased opening of industries from 1 June onwards.
      \begin{itemize}
         \item Workplaces function at 5\% during the period of May 18--31, at 20\% for the month of June, 33\% for the month of July and 50\% from August onwards.

         \item Workplaces function at 5\% during the period of May 18--31, at 33\% for the month of June, 50\% for the month of July and 66.7\% from August onwards.
      \end{itemize}

  \item {\bf Modeling of hotspots and containment zones:} The MHA Order of 15 April 2020 \cite{20200415-MHAOrder} also indicated operational guidelines in hot-spots and containment zones. Hot-spots are areas of large COVID-19 outbreaks, or clusters with significant spread of COVID-19. These are to be determined as per the Ministry of Health and GoI guidelines. In these hot-spots  (red and orange zones in the MHA Order of 17 May 2020 \cite{20200517-MHAOrder}), containment zones may be demarcated. Within these containment zones, there will be strict perimeter control to ensure that no unchecked inward or outward movement of population is permitted, except for maintaining essential services and government activity.

      We generalize and adopt a soft modeling of the containment zones at the level of a ward based on the fraction of hospitalizations observed in that ward. As the fraction of hospitalizations in the ward increases, the ward is more strictly closed with maximum containment being observed when the hospitalized fraction  (of the ward population) in the ward exceeds a configurable parameter; this parameter is currently set to 0.1\%.

  \item {\bf Phased opening of trains:} Suburban trains were not under operation in Mumbai till 31 May 2020. In a large metropolis like Mumbai, trains are a key mode of daily commute to work and even a partial opening of workplaces is infeasible without a corresponding partial opening of the trains. In fact, opening industries without the opening of trains might lead to overcrowding in buses. We simulate the opening of trains in a phased manner, similar to the opening of workplaces.

      We would like to add the caveat that the train transmission parameter, unlike the other parameters of our model, is not calibrated with respect to the initial data prior to the lockdown. We pre-suppose that initial spread of the disease was primarily not via people traveling by trains in Mumbai. Furthermore, trains had been running below capacity a week prior to the lockdown and have stopped running entirely since then. Our choice of the contact rates on trains is based on, on the face of it, reasonable assumptions but, admittedly, not on data. This needs to be fine-tuned over a few weeks, to arrive at better projections and at better strategies for opening up further. A cautious start with restricted occupancies may provide us with more data on the adverse impact of infection spread on trains.

  \item {\bf Case Isolation and Home Quarantine:} We model case isolation and home quarantining of individuals displaying symptoms as follows. After 24 hours of a person showing symptoms, the person is case-isolated and the members of his household are quarantined, if the household is compliant. This implies the person no longer goes to work, the household members remains at home, and the visits to the community centers are lowered significantly. Quarantining of other close contacts arising from contact tracing, e.g., close contacts at work or in the immediate neighbourhood, is not modelled in this work, but will be done in a subsequent work.

  \item {\bf Role of Masks:} The MHA order of 15 April \cite[Annexure~1]{20200415-MHAOrder} made the wearing of masks in public places compulsory. This has been re-emphasized in the MHA order of 30 May 2020 \cite{20200530-MHAOrder}. Various studies suggest that the use of masks considerably reduces the transmission of respiratory virus diseases in workplaces, community and transport spaces \cite{mask-respiratory2009,maskuse-influenza2010,mask-respiratoryvirus2020,chu2020physical}. Face masks may help in three significant ways. They may directly reduce the spread of the respiratory pathogen. They may help reduce the inhalation of the respiratory pathogen. They may also reduce hand-to-face contact. We model the effectiveness of masks by reducing the ability of an infectious individual to transmit the infection by 20\%, and by reducing the chances of a susceptible individual contracting the infection by 20\%, if a mask is worn.

  \item {\bf Compliance levels:} It is upfront unclear as to the extent to which orders on case isolation, home quarantines and masks are complied with by the population. In fact mobility results~\cite{mobilityreport-Google2020}  suggest that there is still significant movement of people. To this end, we bring a compliance factor into our simulation. Even under the strictest of lockdowns, compliance is lower in high-density areas where individuals do not have the luxury of many individual facilities, e.g., toilets, and are forced to access shared facilities. We assume 40\% compliance to lockdown restrictions in such high-density areas and 60\% in other areas. We also compare these results with increased levels of compliance.
\end{enumerate}

\section{Simulation Results}

As alluded to above, we consider two different scenarios of phased opening of workplaces in
the post lockdown phase from 1 June 2020 onwards.

\begin{itemize}
\item
The workplaces function at 5\% for the period of May 18 -- 31,
at 20\% for the month of June, 33\% for the month of July and 50\% for
August onwards.

\item The workplaces function at 5\% for the period of May 18 -- 31,
at 33\% for the month of June, 50\% for the month of July and 66.7\% for
August onwards.
\end{itemize}

These two scenarios are considered under 40\% compliance in
high-density areas and 60\% compliance in low-density areas. This
level of compliance, with additional measures such as containment zones and active mask usage, matches the observed data on fatalities.  We also
consider these scenarios under increased compliance after the lockdown
to highlight the improvements that may be achieved with better
enforcement. As mentioned above, we assume strict case-isolation and
home-quarantine post lockdown, containment zones, use of masks, and restrictions on those
above 65 to stay at home. We present results both when the
trains are operational, and when they are not. The latter is
unrealistic in Mumbai at higher levels of workplace functioning. It is
added to illustrate the impact of trains on the spread of infections.

We first present the plots representing our simulation results, draw
some observations based on these results and then state the modeling
assumptions behind these simulations.
As in Report~\cite{report1}, in our current simulations, two synthetic cities are created that match the aggregate Mumbai demographic
data. For each synthetic city, five independent simulation runs are conducted. The results reported are
the average of these ten  runs.

\def\mycaption{Varying attendance profiles, with trains on and off}
\newcommand{\putfigures}[3]{
  \begin{figure}
    \ContinuedFloat
    \begin{subfigure}[h]{\textwidth}
      \centering
      \includegraphics[width=\linewidth]{#1.png}

      \includegraphics[width=\linewidth]{#1_log.png}
      \caption{#2} \label{#3}
    \end{subfigure}%
    \caption{\mycaption}
  \end{figure}
}

{
  \begin{figure}
    \begin{subfigure}[h]{\textwidth}
      \centering
      \includegraphics[width=\linewidth]{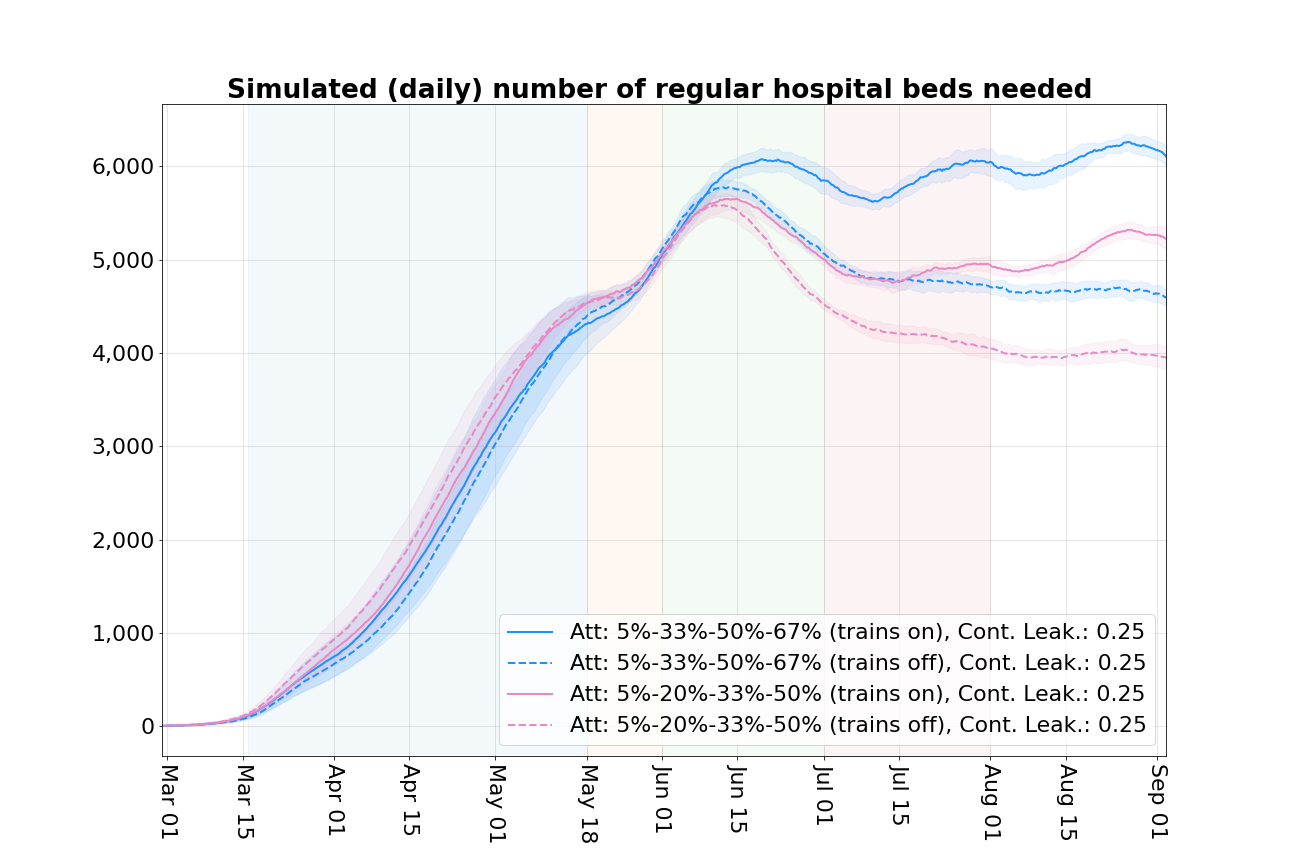}

      \includegraphics[width=\linewidth]{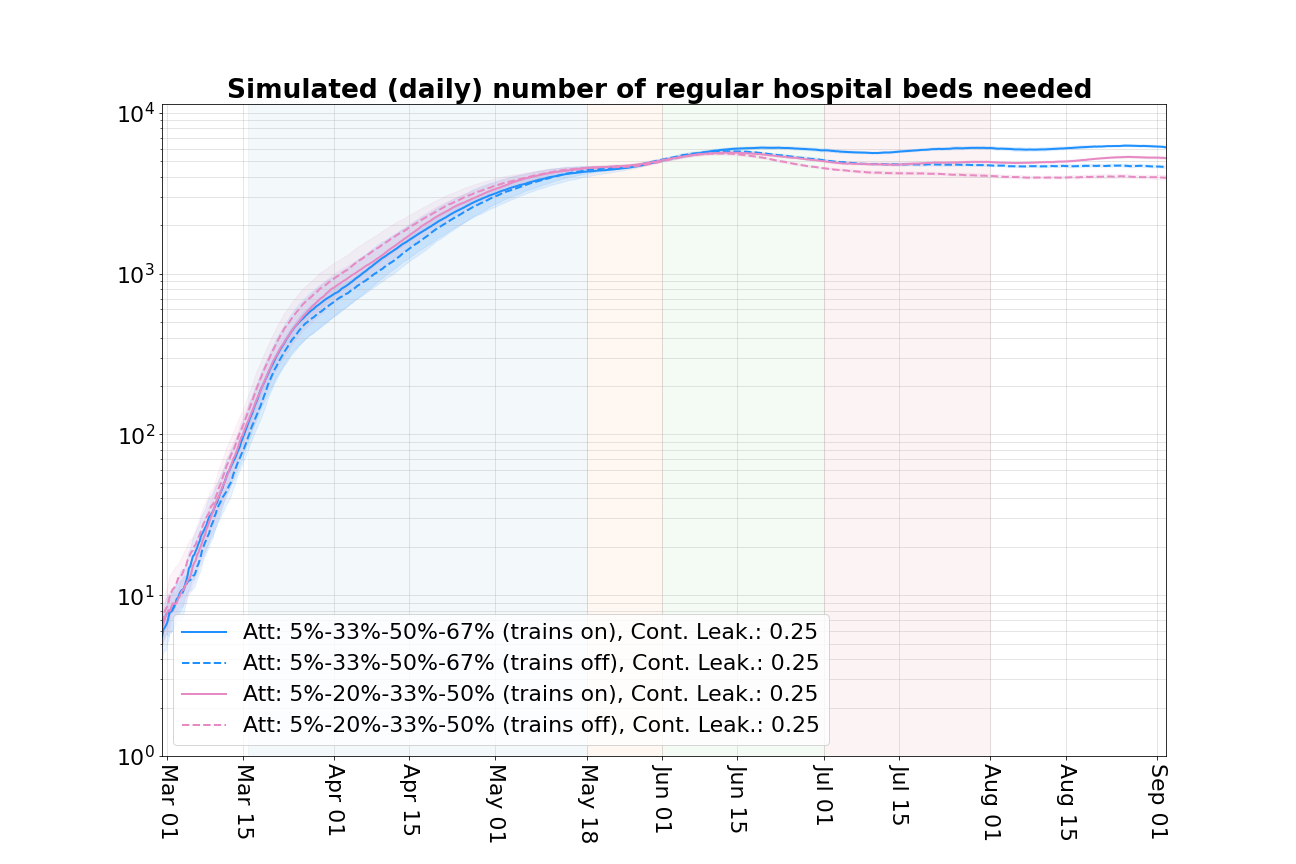}
      \caption{Simulated number of  daily hospitalized patients in linear and log scale.} \label{figure-hospitalized}
    \end{subfigure}%
    \caption{\mycaption}
  \end{figure}

  \putfigures{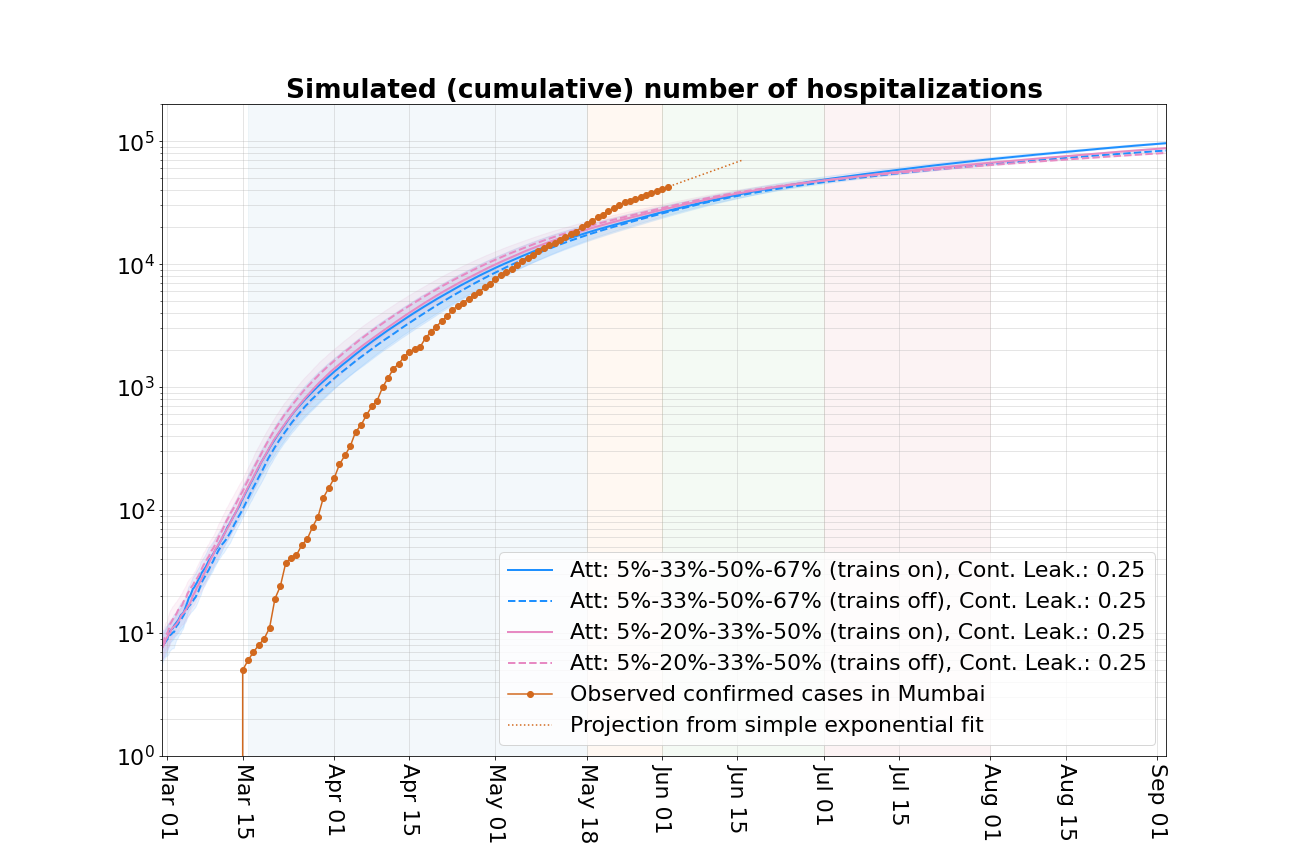}{Simulated number of  cumulative  hospitalized patients in linear and log scale along with the confirmed number of positive cases in Mumbai.}{figure-cumhospitalized}

  \putfigures{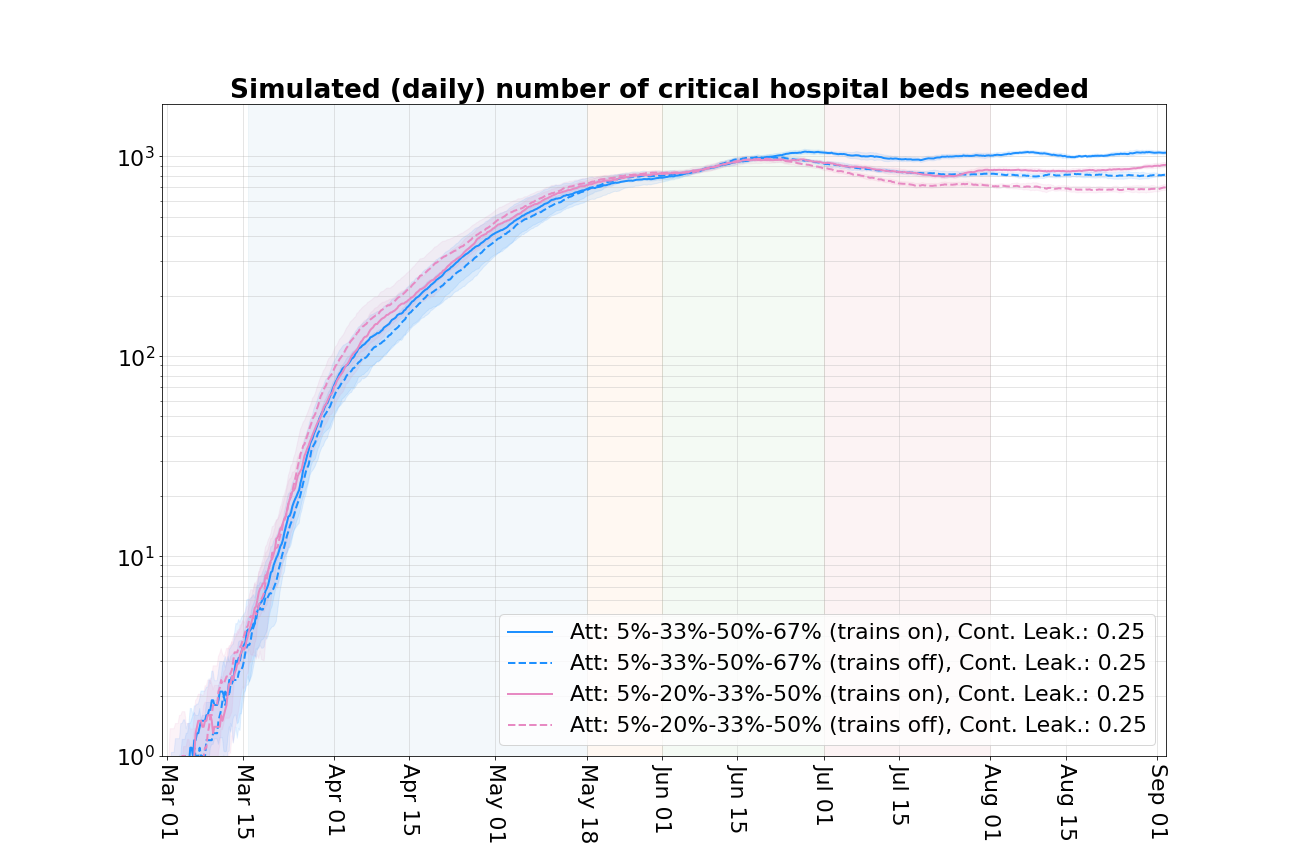}{Number of daily critical cases in linear and log scales.}{figure-critical}

  \putfigures{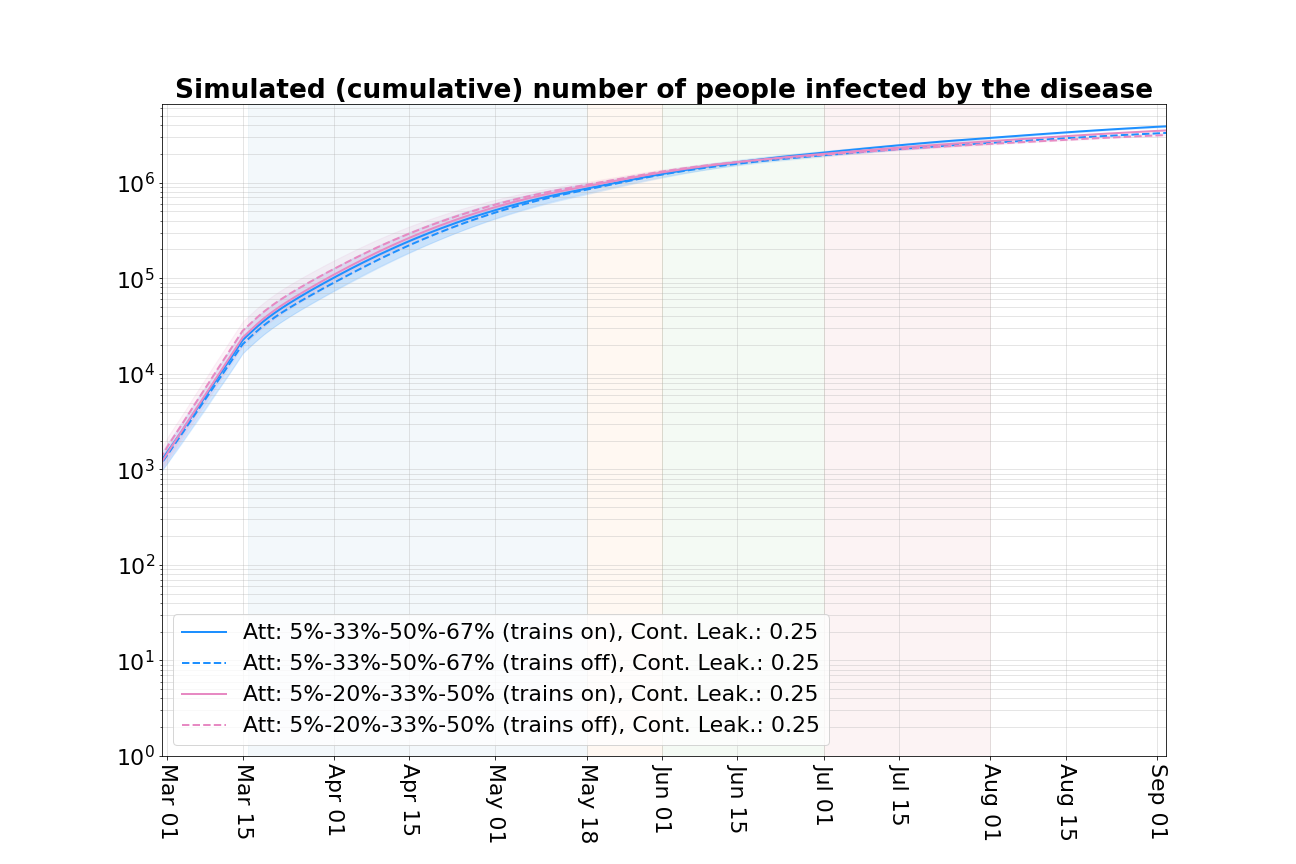}{Cumulative number of people infected by the disease on a linear and log scale.}{figure-affected}

  \putfigures{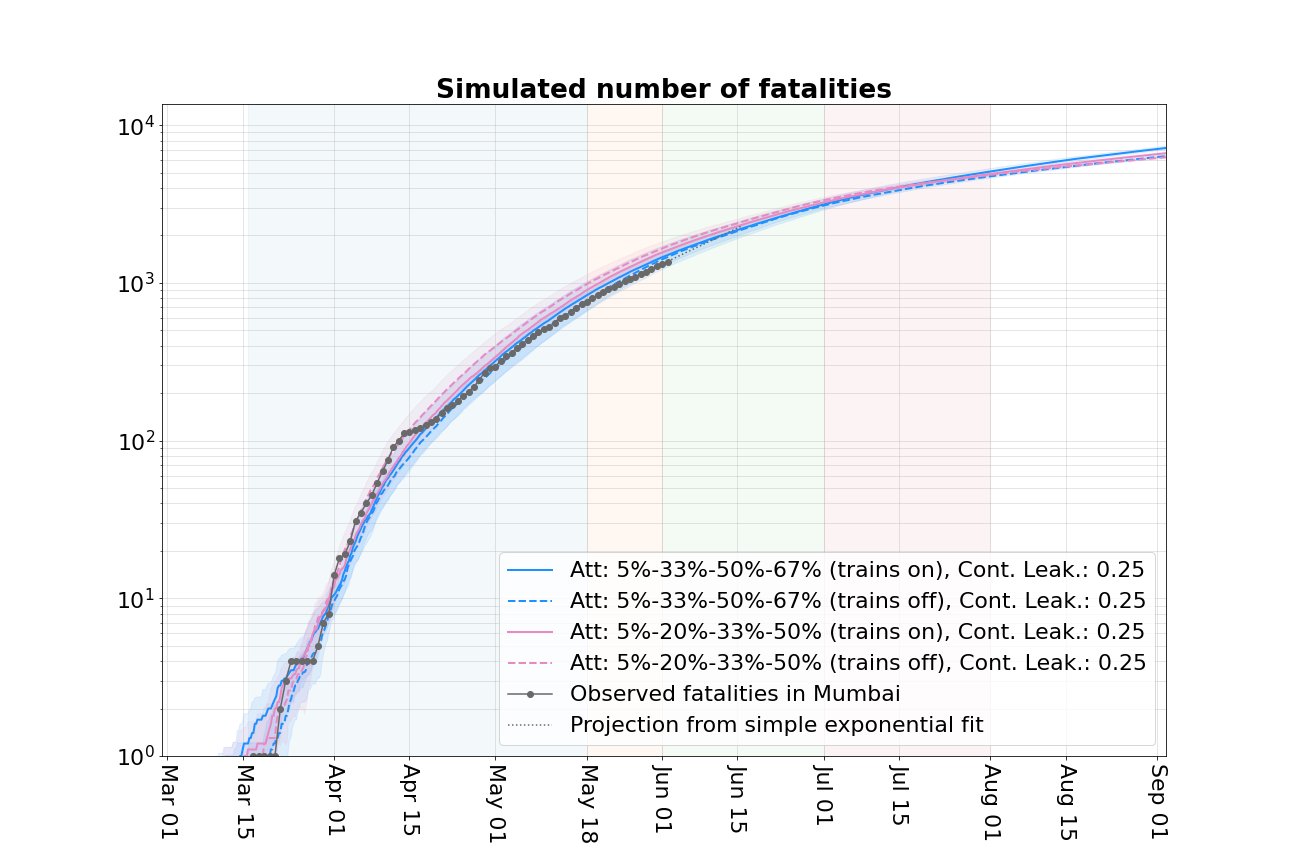}{Cumulative fatality numbers on a linear and log scale along with the observed number of fatalities in Mumbai.}{figure-fatal}

  \begin{figure}
    \ContinuedFloat
    \centering
    \begin{subfigure}[h]{\textwidth}
      \includegraphics[width=\linewidth]{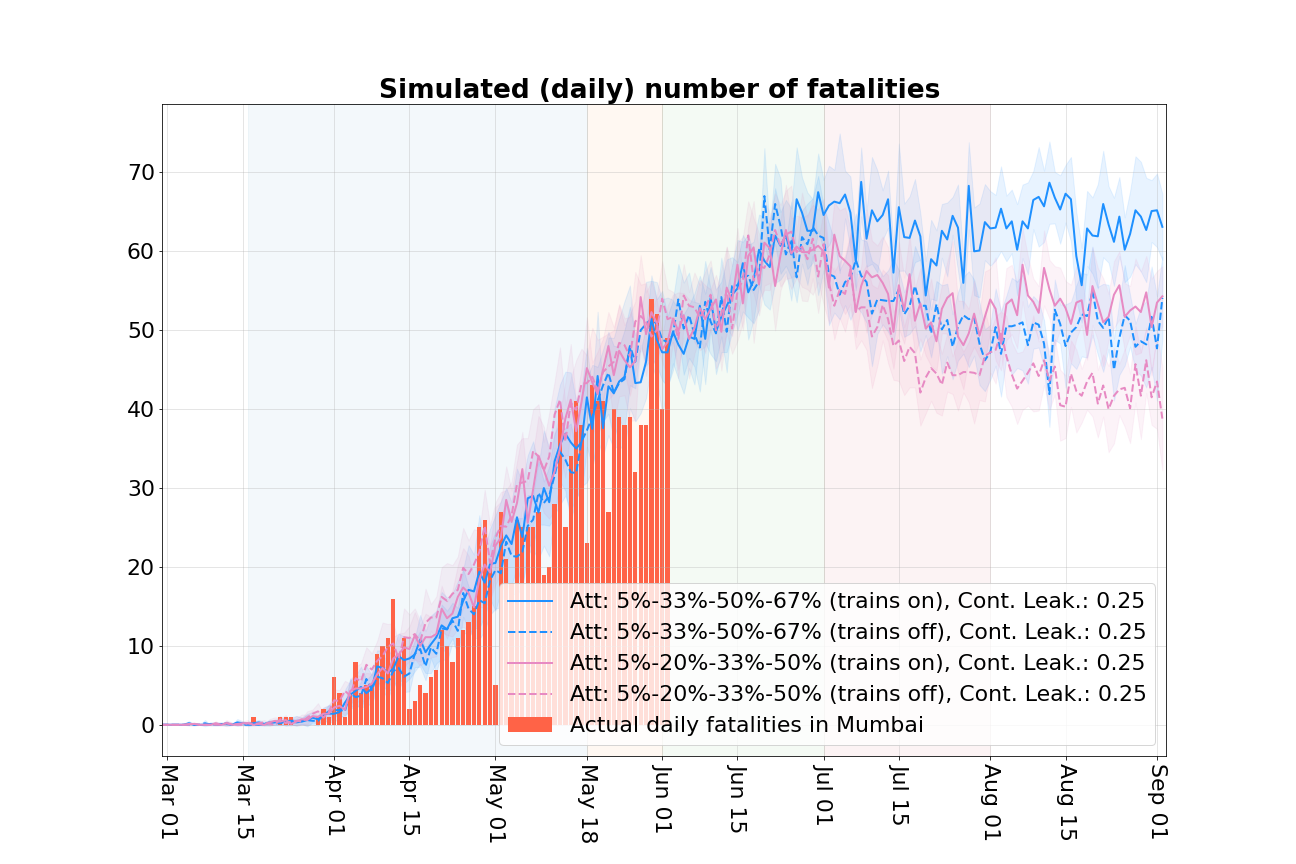}
      \caption{Simulated number of daily fatalities along with the actual
        daily statistics} \label{figure-dailyfatal}
    \end{subfigure}
    \caption{\mycaption}
  \end{figure}
}

{
  \def\mycaption{Varying containment leakages, and compliance (from 1 June).}
  \begin{figure}
    \begin{subfigure}[h]{\textwidth}
      \centering
      \includegraphics[width=\linewidth]{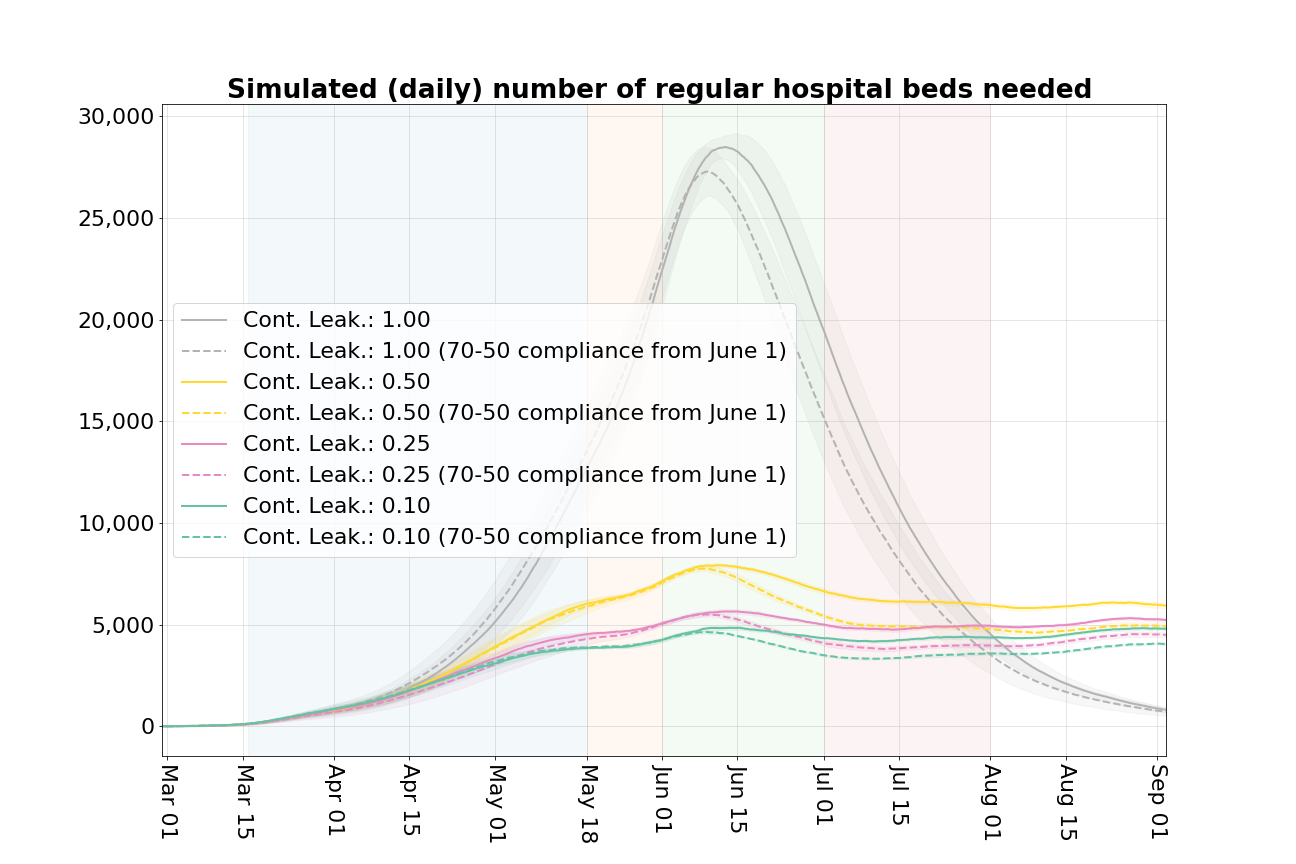}

      \includegraphics[width=\linewidth]{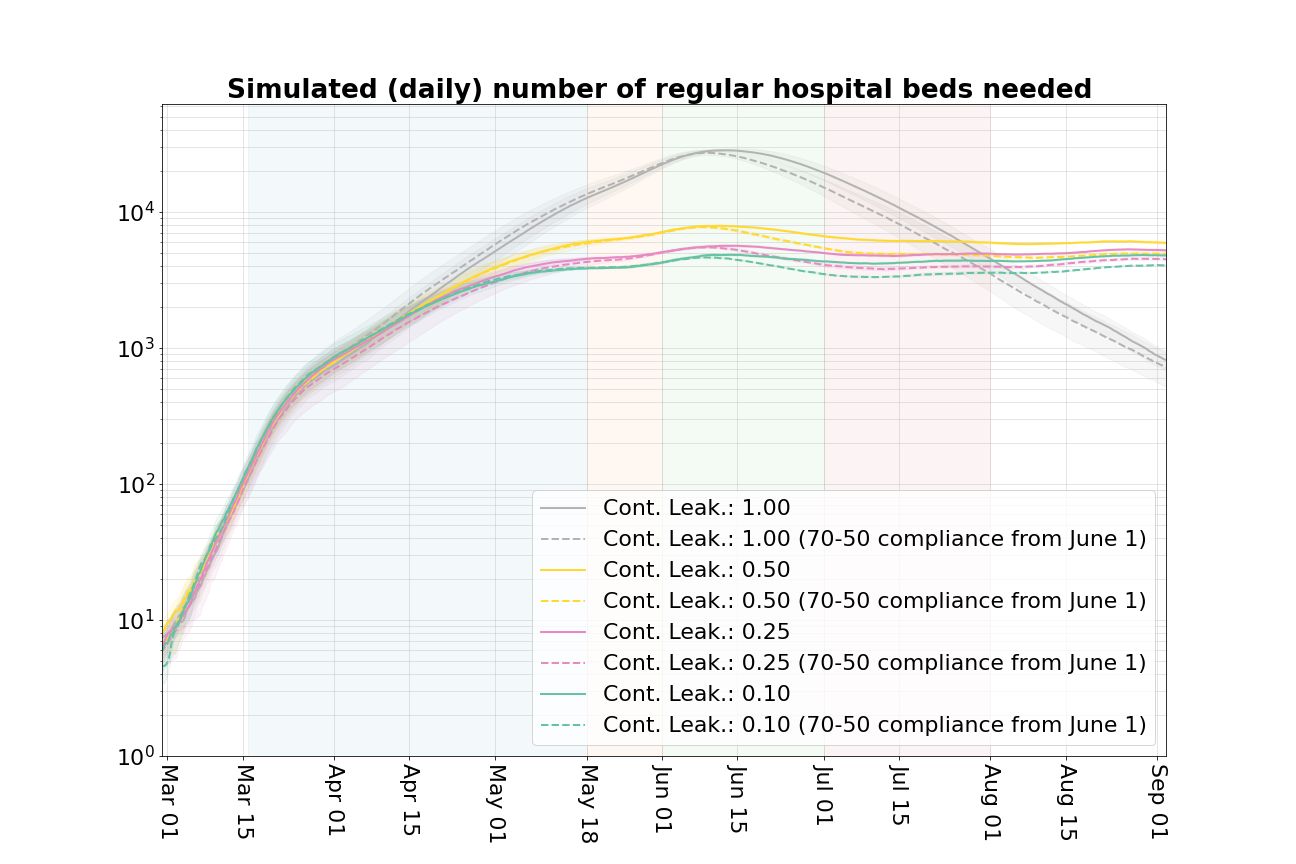}
      \caption{Simulated number of  daily hospitalized patients in linear and log scale.} \label{figure-hosp-cl}
    \end{subfigure}%
    \caption{\mycaption}
  \end{figure}

  \putfigures{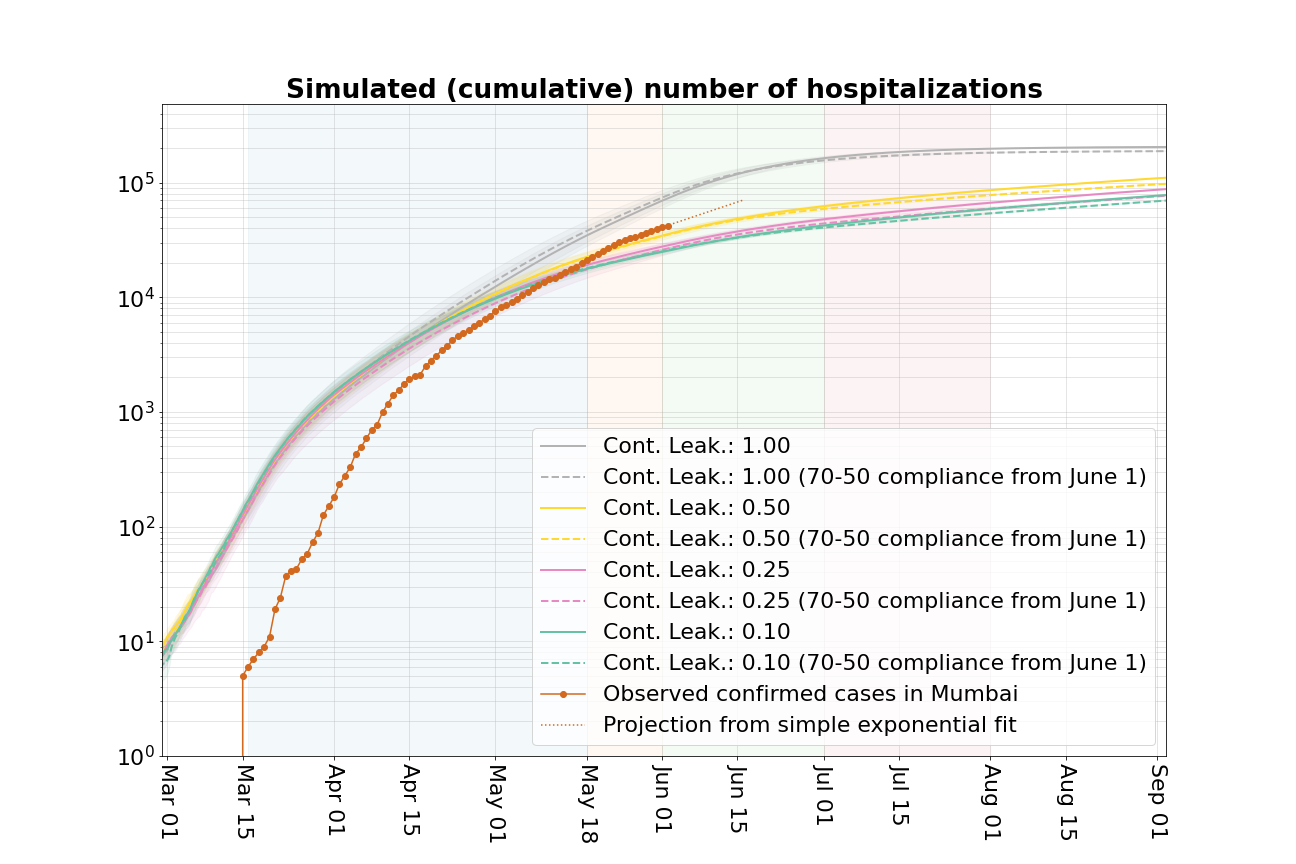}{Simulated number of  cumulative  hospitalized patients in linear and log scale along with the confirmed number of positive cases in Mumbai.}{figure-cumhospitalized-cl}

  \putfigures{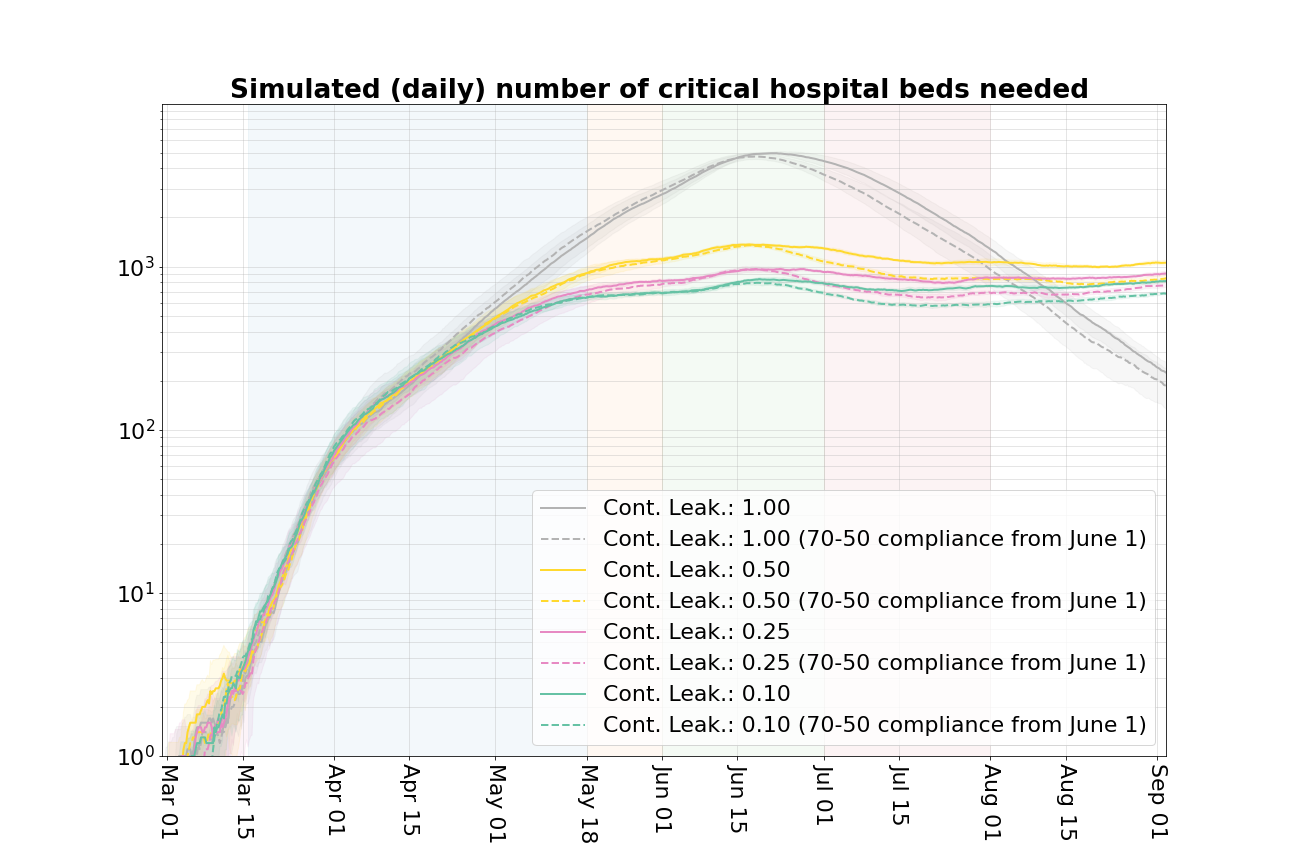}{Number of daily critical cases in linear and log scales.}{figure-critical-cl}

  \putfigures{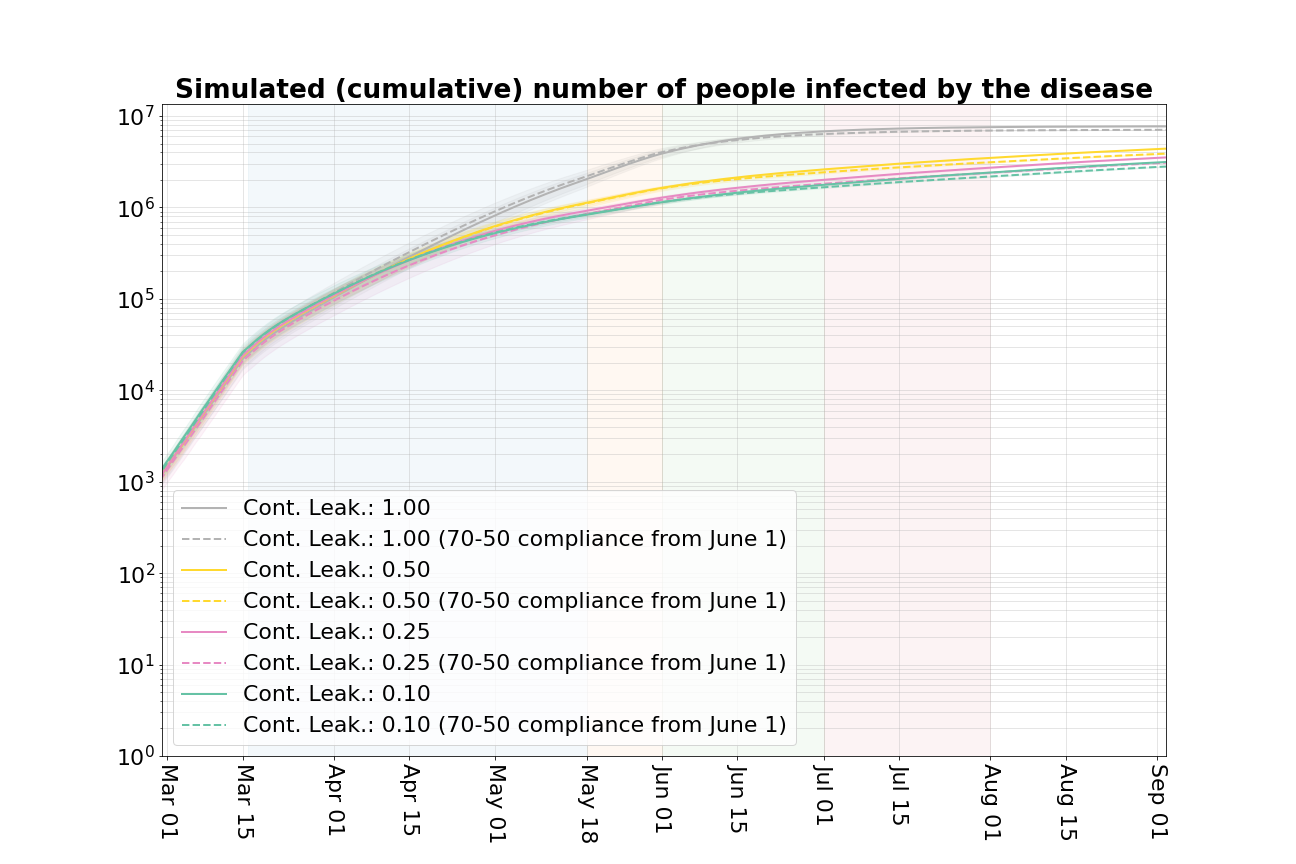}{Cumulative number of people infected by the disease on a linear and log scale.}{figure-affected-cl}

  \putfigures{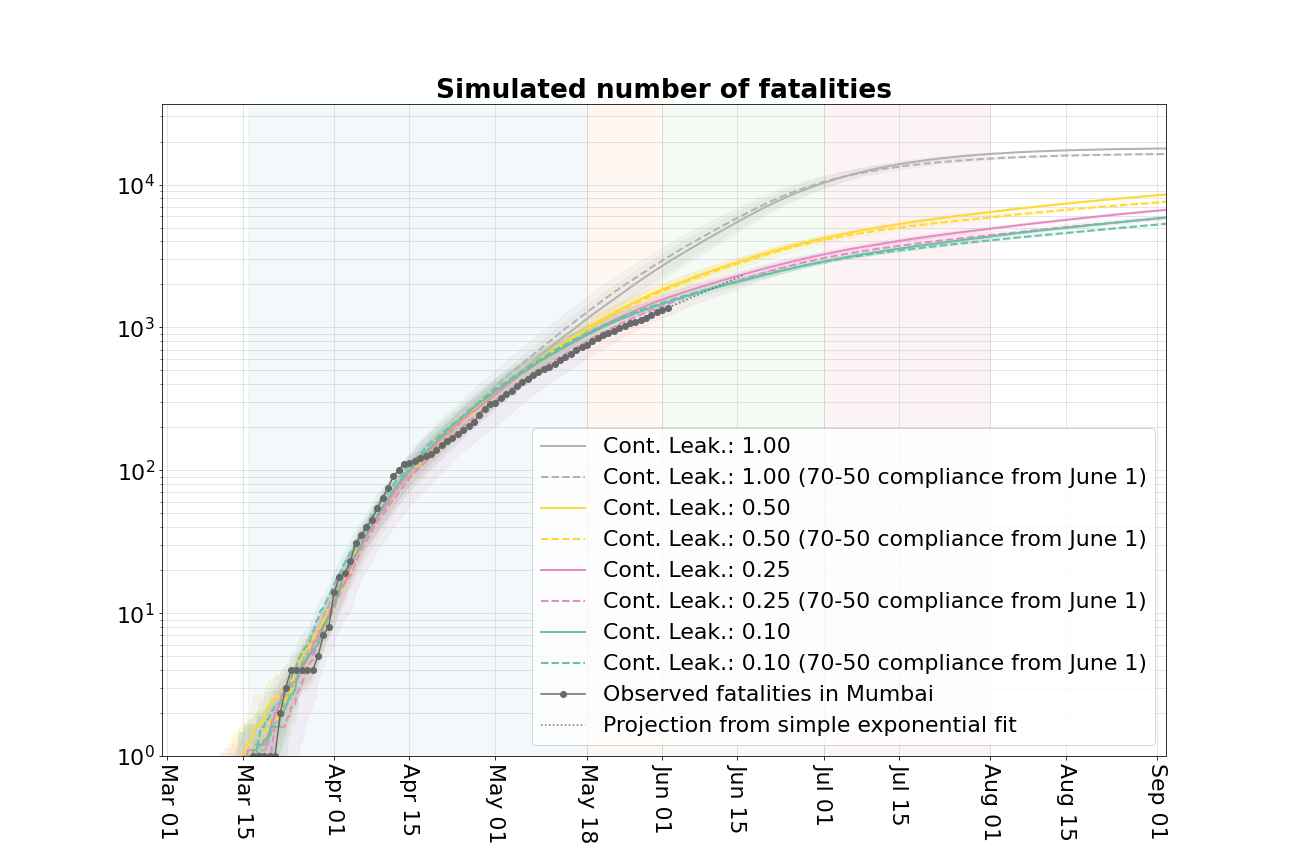}{Cumulative fatality numbers on a linear and log scale along with the observed number of fatalities in Mumbai.}{figure-fatal-cl}
  \begin{figure}
    \ContinuedFloat
    \begin{subfigure}[h]{\textwidth}
      \centering
      \includegraphics[width=\linewidth]{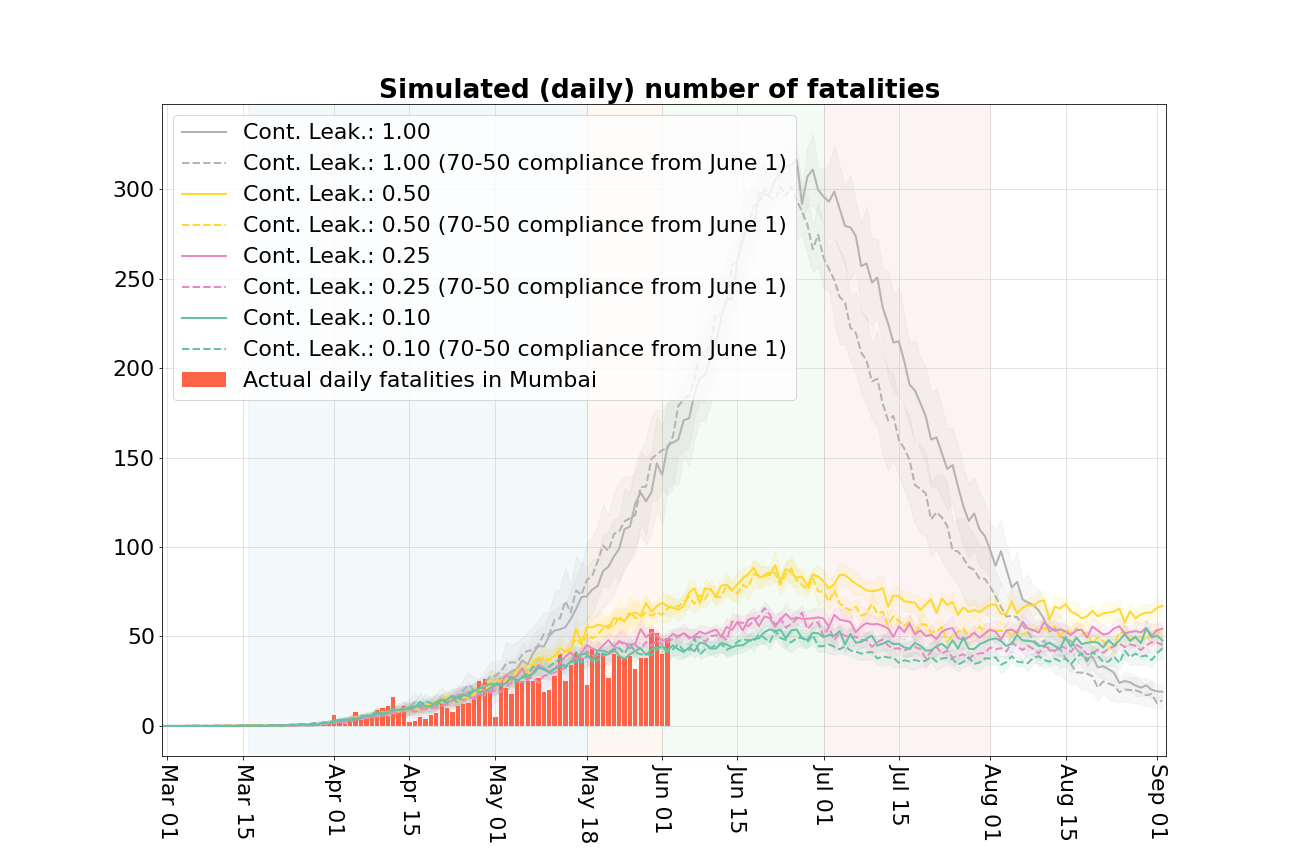}
      \caption{Simulated number of daily fatalities along with the actual
        daily statistics} \label{figure-dailyfatal-cl}
    \end{subfigure}
    \caption{\mycaption}
  \end{figure}
}

In Figure~\ref{figure-hospitalized}, we show the number of
hospitalized patients as a function of the date as per the
simulations while Figure~\ref{figure-cumhospitalized} shows the
cumulative number of hospitalized patients over the same period. These suggest that the city may need of the order of
6,000--7,000 beds from mid-June onwards. The requirement becomes critical
if the trains are started after 31 May as they lead to a significant
increase in infections.

It is noteworthy that the India
hospitalization case definition includes hospitalization for case
isolation. This number heavily depends on the testing protocol
(positive cases and their contacts). This will, therefore, be different
from the actual number of cases or the actual number of seriously ill cases requiring essential hospitalization. In
Figures~\ref{figure-hospitalized}--\ref{figure-cumhospitalized}, we attempt to capture only those
requiring medical intervention associated with essential hospitalization.

Figure~\ref{figure-critical} shows critical cases as predicted by the
simulation. The numbers suggest that critical care units of order 1,500 would be adequate for Mumbai during the months of June, July and August.  In Figure~\ref{figure-affected},
we plot the cumulative number of people infected by the disease. This
includes all people infected with the disease (asymptomatic,
symptomatic, hospitalized, critical). Figures~\ref{figure-fatal}--\ref{figure-dailyfatal} show the
cumulative fatality numbers and daily fatality numbers as a function
of date under various scenarios.

Another point to highlight is the oscillating behaviour in the number of daily hospitalized and critical cases in Figures~\ref{figure-hospitalized} and \ref{figure-critical}. This comes from our specific modelling of containment zones. As the number of hospitalized cases in a ward increases, a partial lockdown in proportion to the number of hospitalized cases comes into play. When the number of hospitalizations exceeds a threshold a full lockdown is enforced. Conversely, as the number hospitalized cases reduces and is below the aforementioned threshold, the partial lockdown is appropriately eased.

To compare between various levels of strictness in this modelling, we now consider a particular phased opening scenario, namely: 5\% for
the rest of May, 20\% for June, 33\% for July and 50\% for August
onwards and vary the containment leakage to see how this affects the
numbers. We plot results for containment leakage of 10\%, 25\%, 50\%
and 100\% (100\% containment leakage corresponds to the case when there
is no adaptive containment enforced in the sense that there are no additional measures enforced as a function of higher hospitalization cases).

Figures~\ref{figure-hosp-cl}--\ref{figure-dailyfatal-cl}
represent the (simulated) number of daily hospitalized, cumulative
hospitalized, daily critical, cumulative fatalities and daily
fatalities respectively for the varying containment leakages, alongside the same with increased compliance of 70\% and 50\% in low density and high-density areas respectively from 01 June 2020. These
plots demonstrate how an effective containment policy (even as low as
50\%) can significantly reduce the number of hospitalized, critical
and fatalities.

\begin{figure}[t]
  \centering
    \begin{subfigure}[h]{0.7\linewidth}
      \centering
      \includegraphics[width=\linewidth]{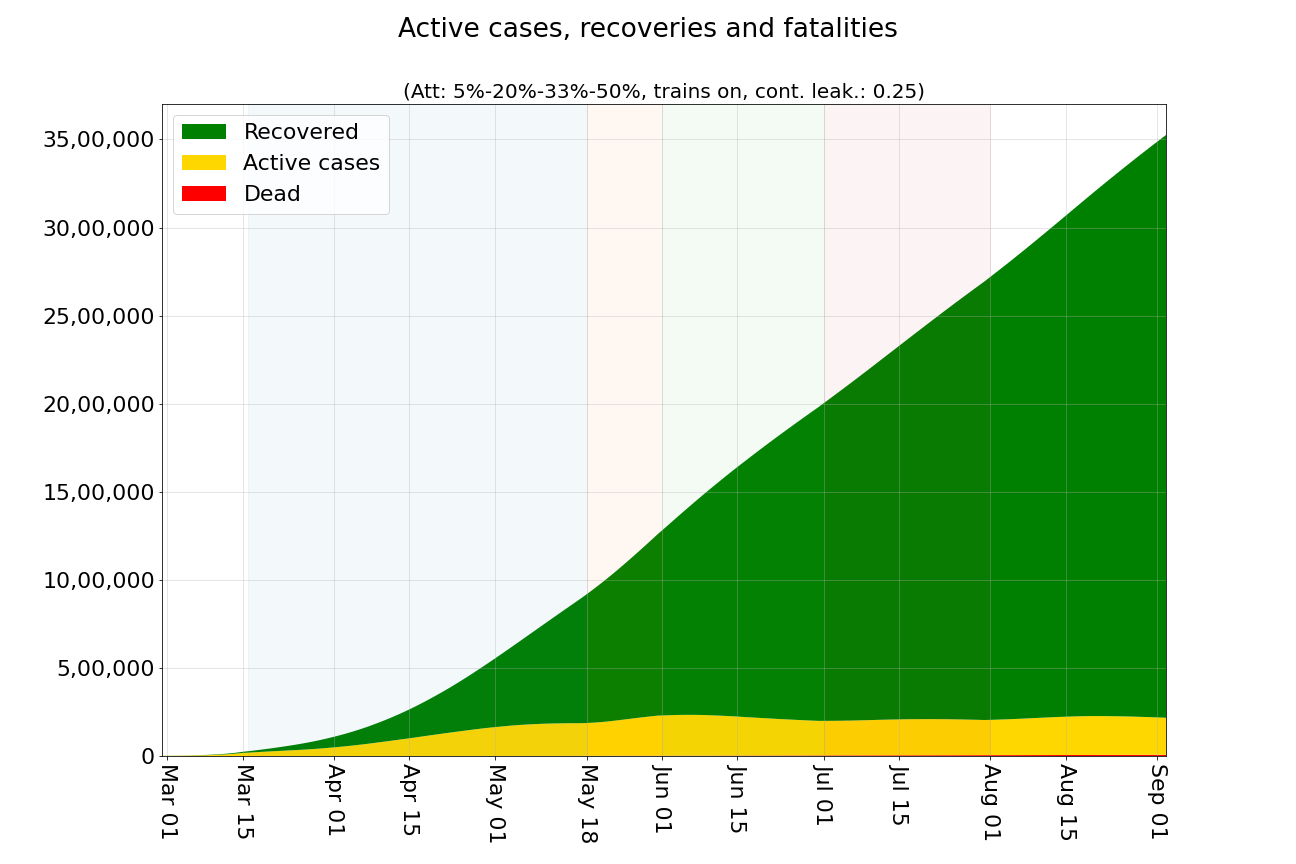}
  \end{subfigure}

  \begin{subfigure}[h]{0.7\linewidth}
    \centering

    \includegraphics[width=\linewidth]{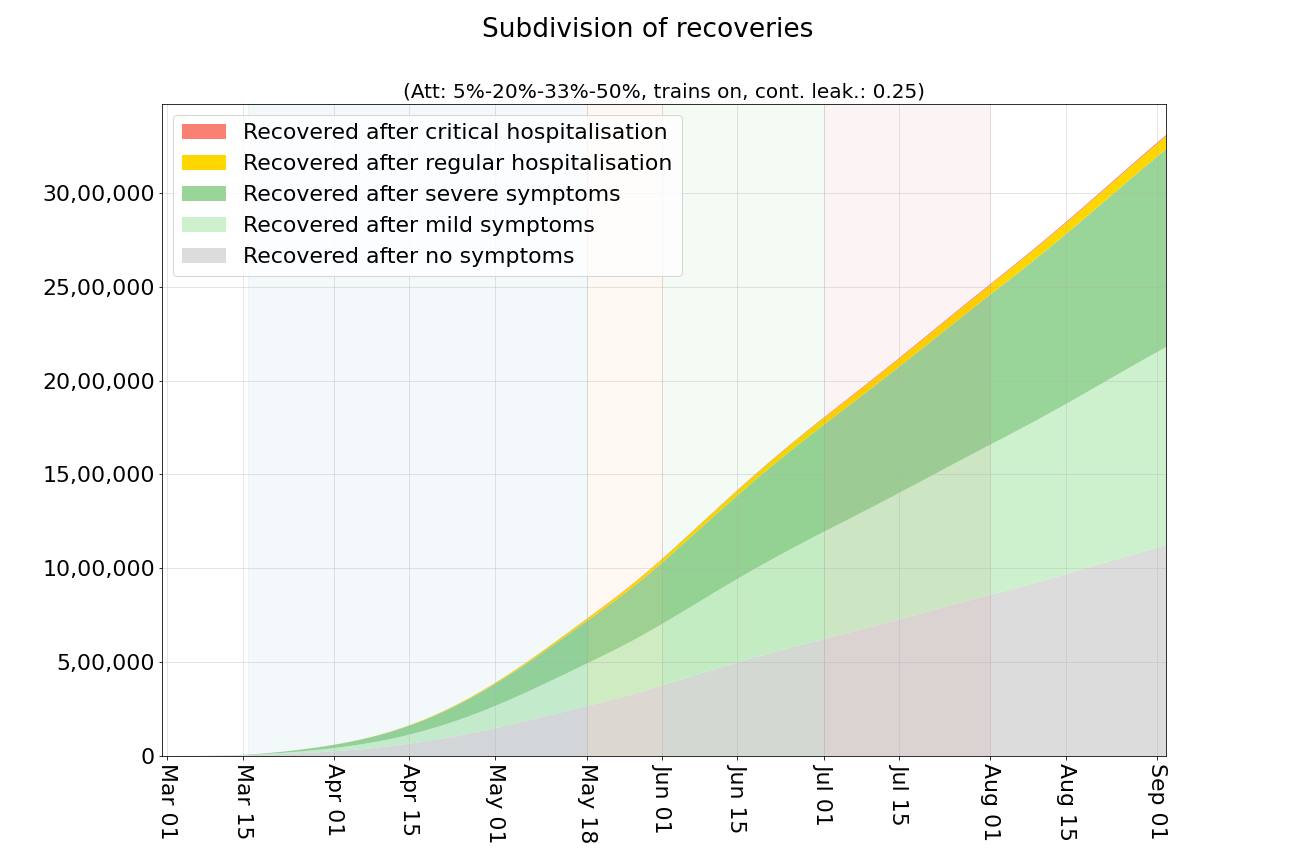}
  \end{subfigure}%

  \begin{subfigure}[h]{0.7\linewidth}
    \centering

    \includegraphics[width=\linewidth]{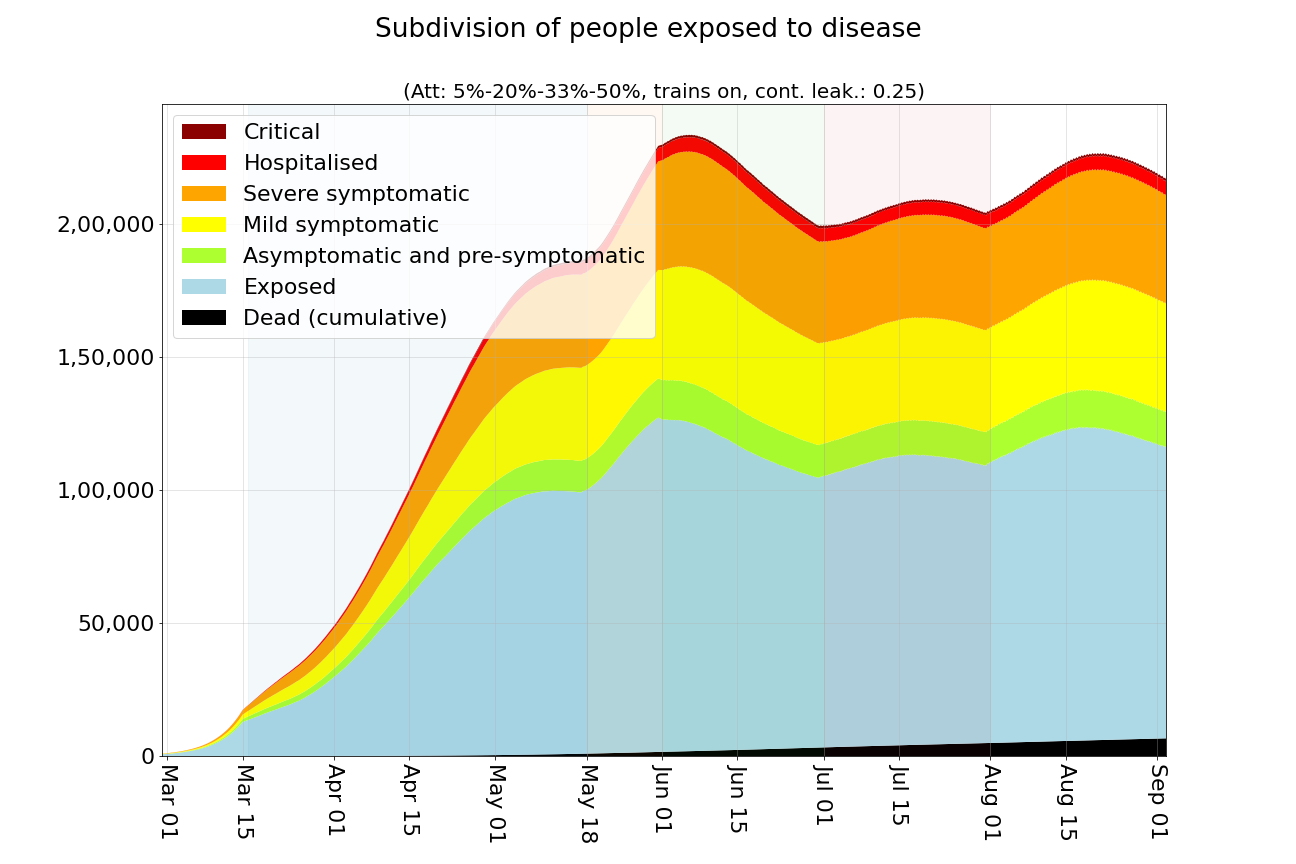}
  \end{subfigure}%
  \caption{Stackplot of division of simulated number infected patients} \label{figure-stackplot}
\end{figure}

Finally, we draw stackplots of the number of
infected people for the particular scenario of 5\% for
the rest of May, 20\% for June, 33\% for July and 50\% for August
onwards, containment leakage of 0.25\% and trains operational in
Figure~\ref{figure-stackplot}. The first figure in
Figure~\ref{figure-stackplot} plots the cumulative number of active cases,
recovered cases and fatalities as a function of time, the second
figure plots a finer daily sub-division of those exposed to the
disease (exposed, asymptomatic, mildly symptomatic, severe
symptomatic, hospitalized, critical and deceased) while the third figure
plots the cumulative sub-division among the recovered cases.

\subsection{Some observations}

\begin{enumerate}
\item
As noted earlier,
the simulation results, conservatively interpreted to account for model uncertainty, suggest that
 about 8,000 beds for COVID-19 patients with severe symptoms requiring hospitalization, and around 1,500 ICU's,
 can manage the medical needs arising from COVID-19 infections in the month of June to August even with partial opening up of workplaces
 and transportation.  Thereafter, continuing stringent controls may keep the numbers at similar levels.
\item
We have assumed asymptomatic infections to be 33\%. Literature suggests that this lies between 20--60\%
\cite{mizumoto2020estimating,ferguson2020report,wang2020evolving}. A lower percentage of asymptomatic infections would imply a higher daily requirement of hospital beds, and a higher percentage would imply a lower daily requirement of hospital beds. Again, it is important to keep in mind that trains may lead to more infections than suggested by our model.
\item
We have used Wuhan based disease progression analysis \cite{ferguson2020report,verity2020estimates}.
Availability of Indian data would greatly help the accuracy of the generated scenarios.
\item
Finally, we end with a caveat that simulation numbers are sensitive to how well the containment zones are managed as well as the compliance levels. Managing these as well as other measures such as social distancing, masks, aggressive contact tracing are critical to keeping the health degradation numbers low.
\end{enumerate}

\section{Modeling Assumptions}
\label{sec:modeling-assumptions}

The assumptions of our model are similar to those in the first
report. We refer the reader to the first report~\cite{report1} for
details on the agent-based model, how the synthetic city is
constructed, the assumptions therein. We state below the additional assumptions to incorporate the
new interventions in this report.

\begin {enumerate}
\item
 \noindent{\bf Case Isolation and Home Quarantine:} After 24 hours of a person showing symptoms, the person and the family
  is set to be quarantined. This implies the person no longer goes to work, the family remains at home, household interactions reduce by 25\%, and the visits to the community centers are lowered significantly to 10\% of the usual.

\item {\noindent \bf Masks:} Assumed to reduce transmission rates at
  workplaces, community and transport by 20\% . This is active from
  April 9th onwards. If an infected person interacts with the
  uninfected person and both are wearing masks, the transmission rate
  reduces to 64\%.  On the other hand, if any one of them is not
  wearing mask, the transmission rate only reduces to 80\%.

\item
{\noindent \bf Containment zones:}
We model containment zones at a ward level. We base the level of containment as an adaptive  function of the active hospitalizations observed in the ward. Number of hospitalizations is taken as the decision variable as it is easily observable as compared to the tracking the number of positive cases in a ward, which may be harder to  estimate accurately without extensive testing. The ward is incrementally closed as more number of hospitalizations are observed in the ward. When the number of hospitalizations in the ward exceeds $0.1\%$, we move to maximum containment -- only $25\%$ of the population is allowed to leave or enter this ward compared to the usual. This $25\%$ includes residents travelling for essential needs, as well as workers involved in essential activities.

Specifically, this percentage of activity (entering and leaving the ward), or containment leakage, is our control and we set the ward accessibility to
\[
\max\left(1 - 7.5 \cdot H,  0.25\right)
\]
where $H$ (in percentage) denotes the fraction of people in the ward that are hospitalized\footnote{In our current implementation, the number of hospitalized cases \emph{excludes} those who are currently in critical care facilities.} (described pictorially in Figure~\ref{fig:containment} with varying levels of leakage parameters). The simulator incorporates the ward accessibility by dampening all interactions in community spaces, and fraction of people who go to work from this ward, by the above accessibility factor.
\begin{figure}[h]
  \begin{center}
    \includegraphics[width=0.75\linewidth]{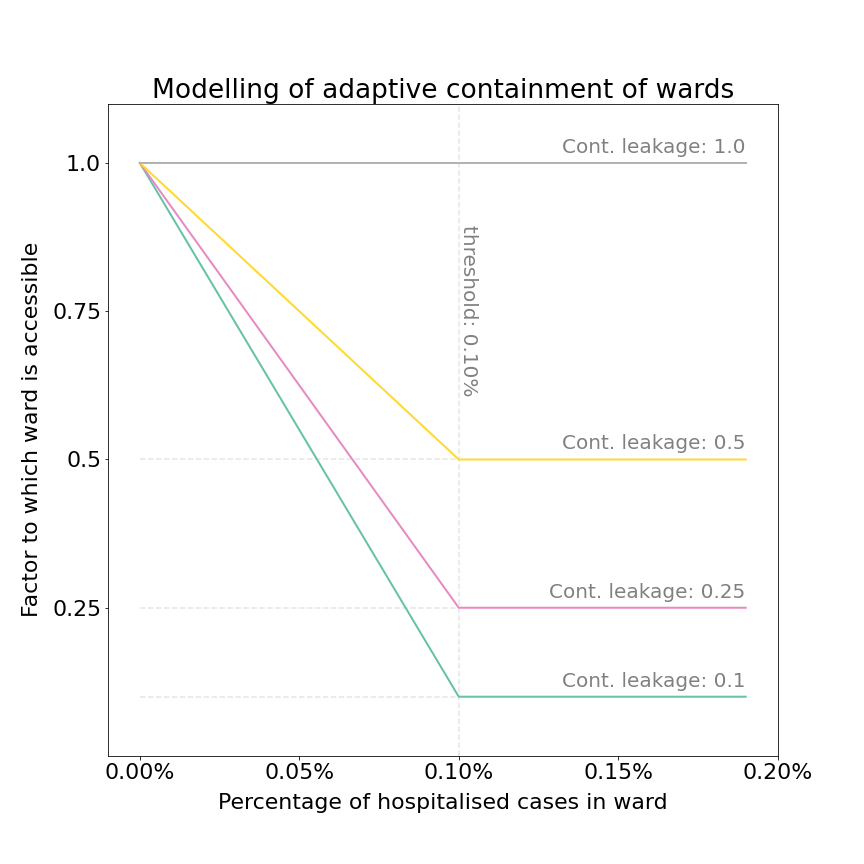}
  \end{center}
  \caption{Adaptive ward containment}
  \label{fig:containment}
\end{figure}

\item
{\noindent \bf Age restrictions:}
We assume that people sixty five years and older are restricted to stay at home.

\item
{\noindent \bf Trains:}
For Mumbai, suburban trains are a key mode of daily transportation with
approximately 78 lakh (or 7.8 million) passenger trips daily in normal
times. However, trains were stopped in Mumbai prior to the national lockdown and
were running below capacity for at least a week before that. Moreover, it is
assumed that the initial infections were seeded by travellers that came from
abroad. The primary mode of travel for this group is unlikely to be suburban
rail transport. Therefore, we do not calibrate the train transmission parameter
to the data prior to the lockdown.

However, as mentioned earlier, all other transmission parameters are calibrated
to data prior to the lockdown.  To estimate the transmission parameter for trains, we
instead use a heuristic argument to compare it to the household transmission
rate.  This is done using assumptions on the relative rates of close contacts
per unit time that an individual might observe at home, as compared to what she
might observe during a commute.  In particular, we assume that the rate of close
contacts that an individual observes in a household is roughly 50/day, while the
same figure during a train commute is roughly 1/minute. The details of this
heuristic calculation follow.


\vspace*{.25cm}

{\em Interaction parameters for trains}:
{
  \newcommand{\ai}{a}
\newcommand{\Ai}{a}
\newcommand{\inter}{close interaction}
\newcommand{\Inter}{Close interaction}

In this section, we spell out our methodology for modeling the suburban train
network in Mumbai.  At time $t$, let $G(t)$ be the set of individuals who are
taking the train (an individual's membership in this set is determined by
factors such as whether the individual is quarantined, whether her office is
closed, etc.).  Let $I(t) \subseteq G(t)$ be the subset of these individuals who
are infectious at time $t$.  We let $d_i$ represent the (one-way) commute
distance of individual $i$.  Let $N$ denote the number of individuals who would
take the train during normal circumstances (e.g, before the epidemic). Define
\begin{align*}
  f_{\text{travel}}(t) &:= \frac{|G(t)|}{N},\\
  D_{\text{infectious}}(t) &:=  \sum_{i \in I(t)}d_i,\\
  D_{\text{total}}(t) &:= \sum_{i \in G(t)} d_i\text{, and}\\
  f_{\text{infectious}}(t) &:= \frac{D_{\text{infectious}}}{D_{\text{total}}}.
\end{align*}
We parameterize the spread of the infection during a suburban rail commute by a
parameter $\beta_T$ with dimensions of (length)${}^{-1}\cdot$(time)${}^{-1}$.
Then, for a susceptible individual $a \in G(t)$, we take the contribution of her
rail commute to her instantaneous infection rate $\lambda_a(t)$ (as defined in
the companion report~\cite{report1}) to be
\begin{equation}
  \beta_T \cdot d_a \cdot  f_{\text{infectious}}(t) \cdot  f_{\text{travel}}(t),\label{eq:3}
\end{equation}
when $t$ represents a time-step in the simulation corresponding to ``morning''
or ``evening'', and $0$ otherwise (we recall from \cite{report1} that each day
is divided into four simulations steps, so that ``morning'' and ``evening'' can
be taken to be the second and the fourth of these four steps).  This is done to
model the fact that peak commute hours are in the mornings and evenings.  The
fraction $f_{\text{travel}}$ represents an attenuation resulting from occupancy
in trains being lower as a result of the intervention being simulated, while the
fraction $f_{\text{infectious}}$ is a heuristic estimate of the chance that a
given \inter{} during the commute is with an individual who is currently
infectious.  Here, \ai{} \emph{\inter{}} is defined as a contact that occurs at
sufficient proximity and for a length of time sufficient for the diseases to be
potentially transmitted.

Ideally, one would like to estimate the value of the parameter $\beta_T$ by a
calibration similar to that used for estimating the other model parameters (such
as $\beta_H$).  However, as outlined in \Cref{sec:modeling-assumptions}, we do
not have sufficient data for doing so.  We therefore try to estimate it
heuristically using the calibrated estimates for $\beta_H$.

Let $c$ represent the probability of \ai{} \inter{} with an infectious individual
leading to a transmission of the infection.  Suppose that the probability law of
the \inter{}s that an individual has in a day with other members in a household
is a Poisson process with rate $r_H$.  Then, the interpretation of the household
interaction parameter $\beta_H$ is as the rate of a process obtained by thinning
this process by a factor $c$:
\begin{equation}
\beta_H = c r_H.\label{eq:1}
\end{equation}
(For further discussion on $\beta_H$, we refer to the accompanying
report~\cite{report1}.)

Now, we model infection spread in trains as follows.  We assume that during
normal operation of the trains (i.e., before the epidemic) during peak commute
hours, an individual would have \inter{}s at the rate $r _T$ during a train
commute, where in a train commute we include last-mile means of transport such
as buses or shared auto-rickshaws/taxis.  The rate parameter governing an
individual getting an infection during commute, if all her \inter{}s were with
infectious individuals, is then
$\rho_T := c \cdot r_T = \frac{r_T}{r_H}\cdot \beta_H$.  (To estimate the actual
rate of infection, $\rho_T$ will need to be attenuated by the factors
$f_{\text{infectious}}$ and $f_{\text{travel}}$ defined above, in order to
account for the fact that not all \inter{}s will be with infectious
individuals.)

Let $s$ be the average speed of commute (including, as above, last-mile means of
transport), so that $\tau_a := d_a/s$ is an estimate of the time of the
(one-way) commute of individual $a$. Thus, during a time step $t$ of the
simulation which includes one direction of the individual's commute, the
contribution of the commute to the instantaneous infection rate $\lambda_a(t)$
of the individual (amortized over the length $\Delta t = 1/4$ day of the
time-step) can be taken to be
\[
  \frac{\tau_a}{\Delta t} \cdot \rho_T \cdot f_{\text{infectious}} \cdot
  f_{\text{travel}} = \frac{d_a \cdot r_T}{\Delta t \cdot s \cdot r_H} \cdot
  \beta_H \cdot f_{\text{infectious}} \cdot f_{\text{travel}}.
\]
Equating the right hand side of this equation with the form for this
contribution posited in \eqref{eq:3}, we therefore get:
\begin{equation}
  \beta_T = \frac{r_T}{s \cdot r_H \cdot \Delta t}\cdot \beta_H.\label{eq:4}
\end{equation}

\paragraph{Heuristic assumptions} We now make the following heuristic
assumptions on the values of the parameters.  We assume, first, that the rate
$r_H$ of \inter{}s in an household for an individual is roughly
{$50 \text{ day}^{-1}$}.  (Assuming an average household size of roughly $5$,
this corresponds to 12.5 \inter{}s per day with each of the other
four members of this average household).

We further assume that during a commute, the rate $r_T$ of close interactions is
roughly $1 \text{ min}^{-1}$ $= 60 \text{ hr}^{-1}$.  Note that \eqref{eq:4}
shows that for estimating $\beta_T$ from $\beta_H$, it is only the ratio
$r_T/r_H$ that is relevant; the actual values of $r_T$ and $r_H$ do not matter
for this calculation so long as this ratio can be estimated well.

\paragraph{Numerical estimates} With the above assumptions, we can now make a
numerical estimate for $\beta_T$ in terms of $\beta_H$, using \eqref{eq:4}.  We
take the average speed of commute (recalling that a commute may include, in
addition to trains, such last-mile means of transport as buses and shared
auto-rickshaws/taxis) as $s = 25 \, \text{km}\, \text{hr}^{-1}$.  With the above
assumptions on the values of $r_H$ and $r_T$, and using
$\Delta t = 1/4\, \text{day}$, this gives
\begin{equation}
  \beta_T = \frac{60 \, \text{hr}^{-1}}{25 \, \text{km} \, \text{hr}^{-1} \cdot 50 \, \text{day}^{-1} \cdot (1/4) \, \text{day}} = 0.192 \, \beta_H\, \text{km}^{-1} \approx \frac{1}{5} \beta_H\, \text{km}^{-1}.\label{eq:2}
\end{equation}
We note here that we represent $\beta_T$ in units of
$\text{km}^{-1}\,\text{day}^{-1}$, $\beta_H$ in units of $\text{day}^{-1}$,
$d_a$ in $\text{km}$, and $\Delta t$ in units of days.

The above calculation depends upon several heuristic assumptions, eg. on the
values of $r_H$ and $r_T$.  While we expect these to have the right order of
magnitudes, we have not, unfortunately, been able to calibrate them against any
directly observed data.  As a caution, we therefore also run our simulations
with the multiplier $\frac{1}{5}$ in \eqref{eq:2} replaced by $1/4$ and $1/6$
respectively.  Some of these results are presented in Figure~\ref{fig:compare-beta-travel} (with higher, base and lower estimates referring
to the multipliers set to $1/4$, $1/5$ and $1/6$ respectively).

A caveat is in order: if the ratio between the rate of close contacts during a
commute, and the rate of close contacts in a household is higher than the value
obtained using the above assumptions, then the effect from infections arising
from train commutes would be higher than what our simulations show.

{
  \def\mycaption{Effect of changing the values of
    $\beta_T$}
  \begin{figure}
    \centering
    \begin{subfigure}[h]{\textwidth}
      \includegraphics[width=\linewidth]{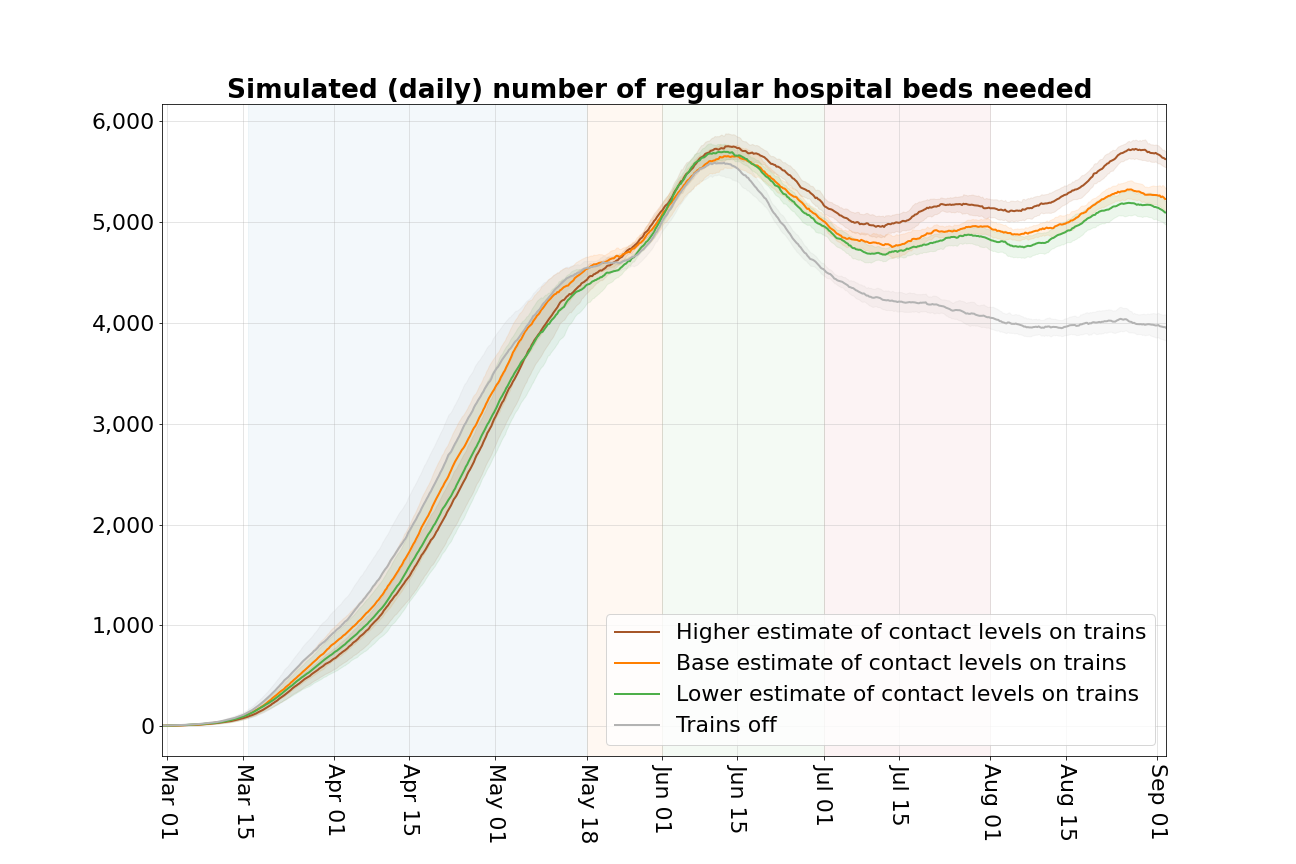}
      \includegraphics[width=\linewidth]{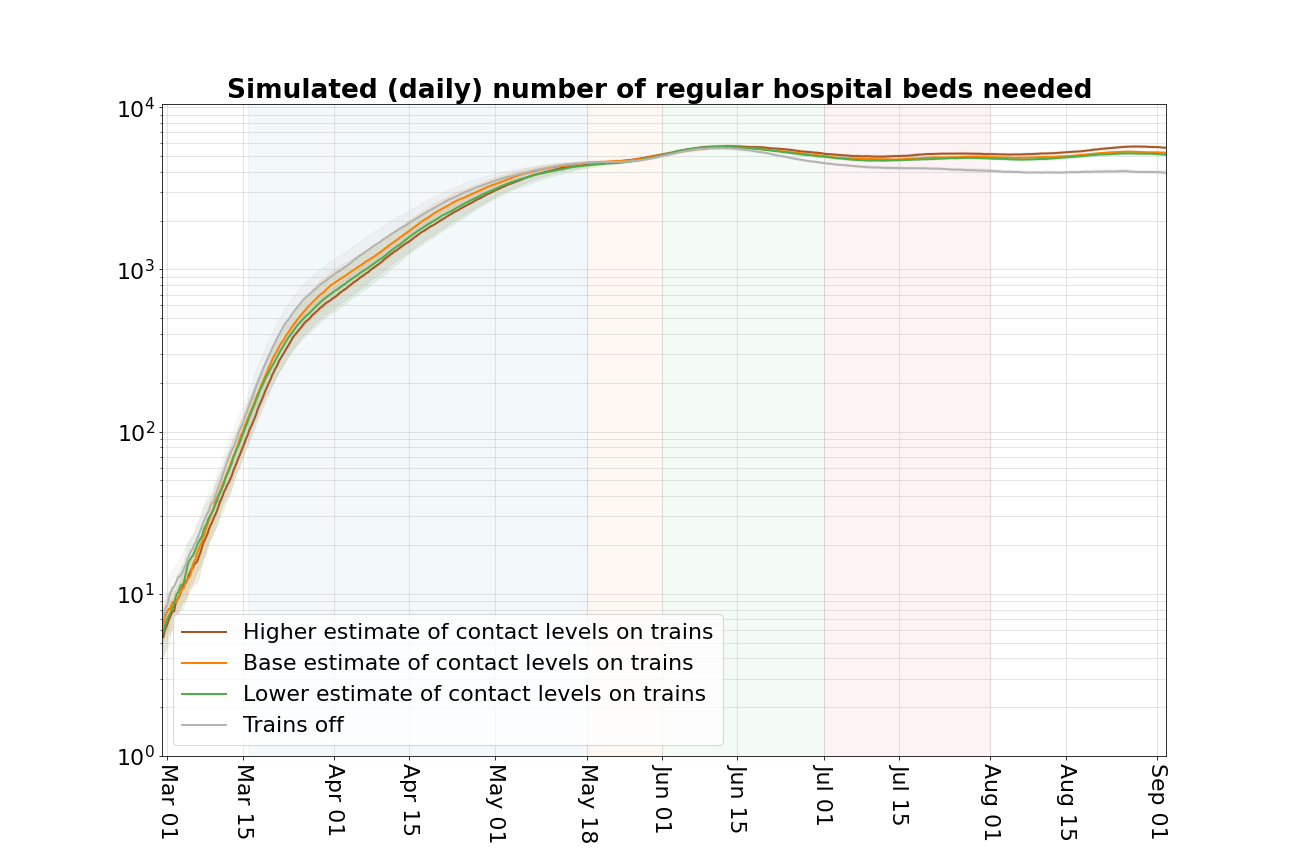}
      \caption{Simulated number of  daily hospitalised patients in linear and log scale.} \label{figure-trains}
    \end{subfigure}%
    \caption{\mycaption}
    \label{fig:compare-beta-travel}
  \end{figure}

  \putfigures{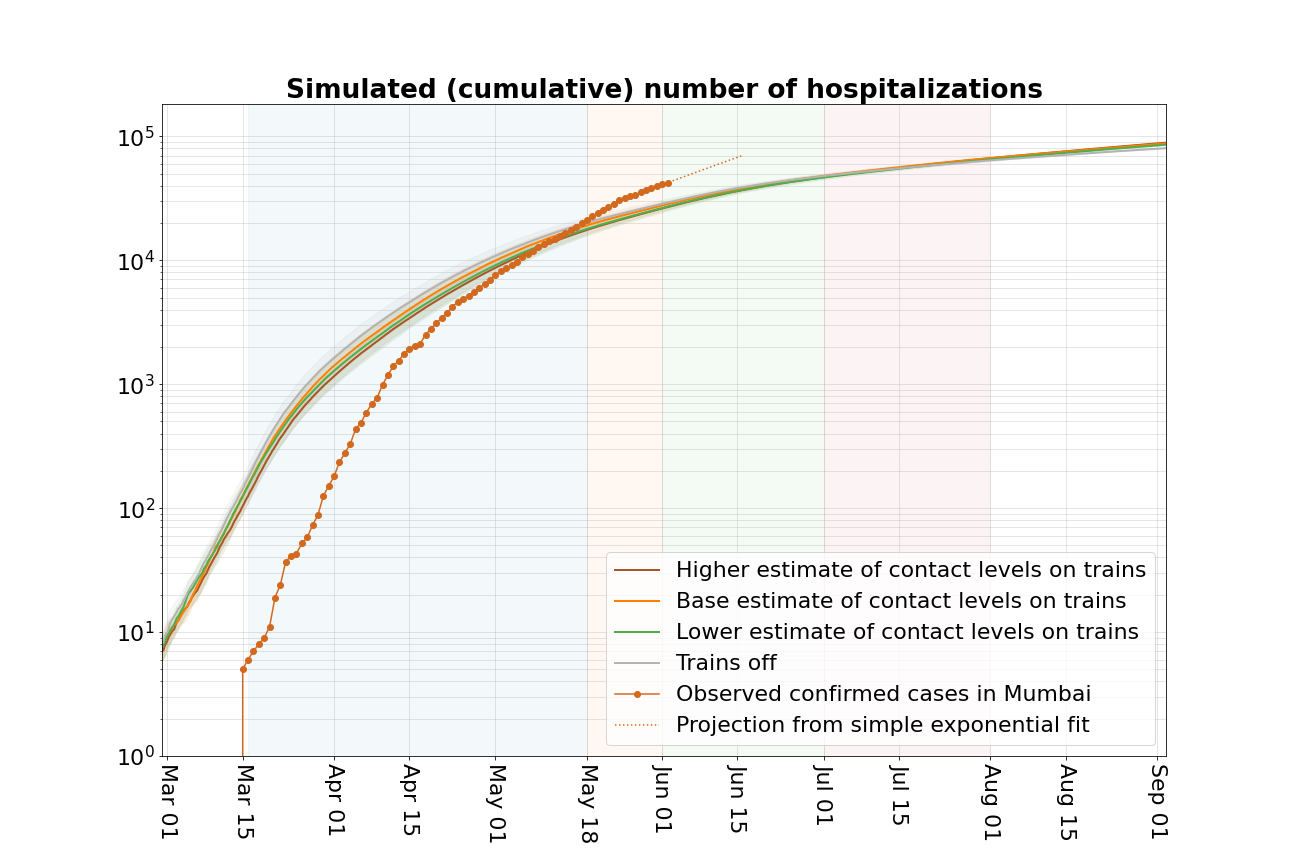}{Simulated number of  cumulative  hospitalised patients in linear and log scale along with the confirmed number of positive cases in Mumbai.}{figure-cumhospitalized-trains}

  \putfigures{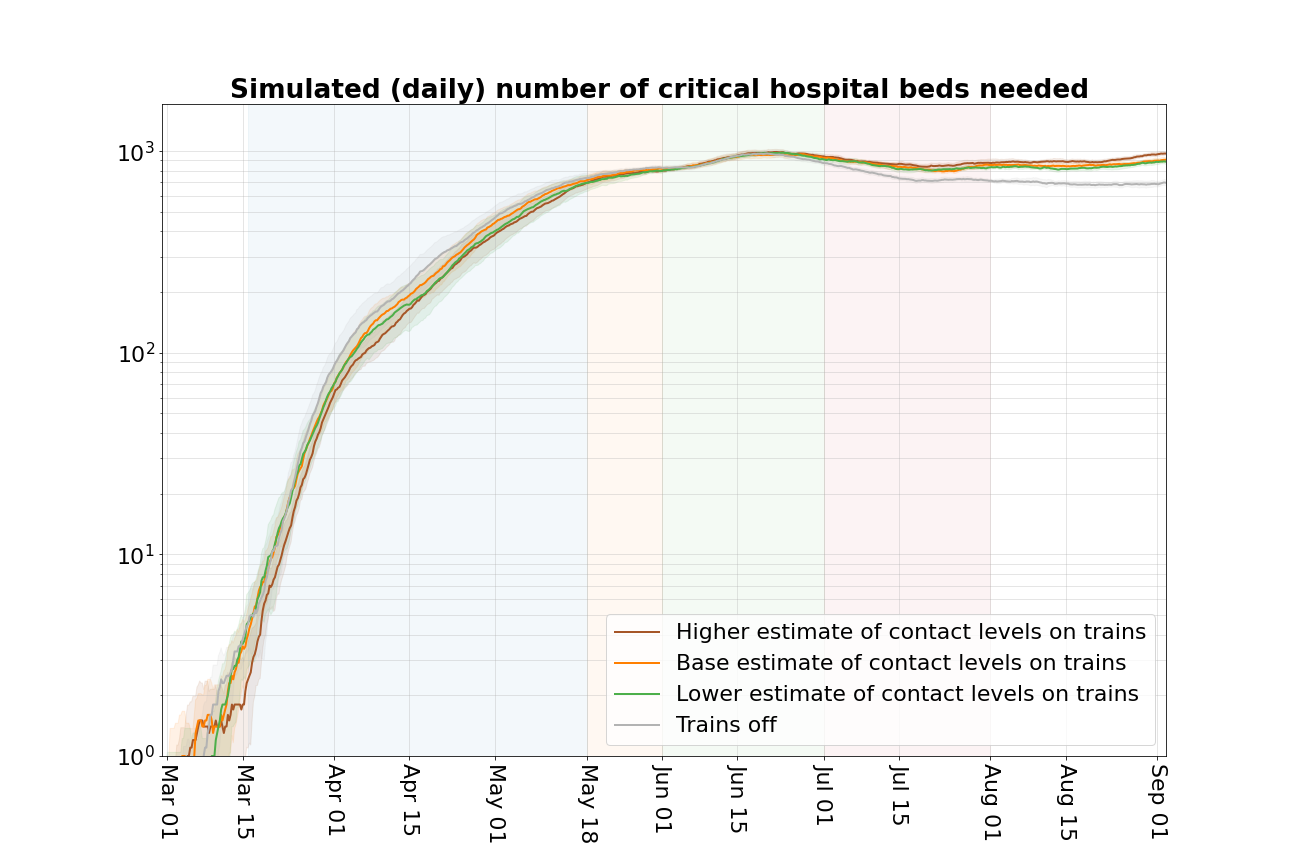}{Number of daily critical cases in linear and log scales.}{figure-critical-trains}

  \putfigures{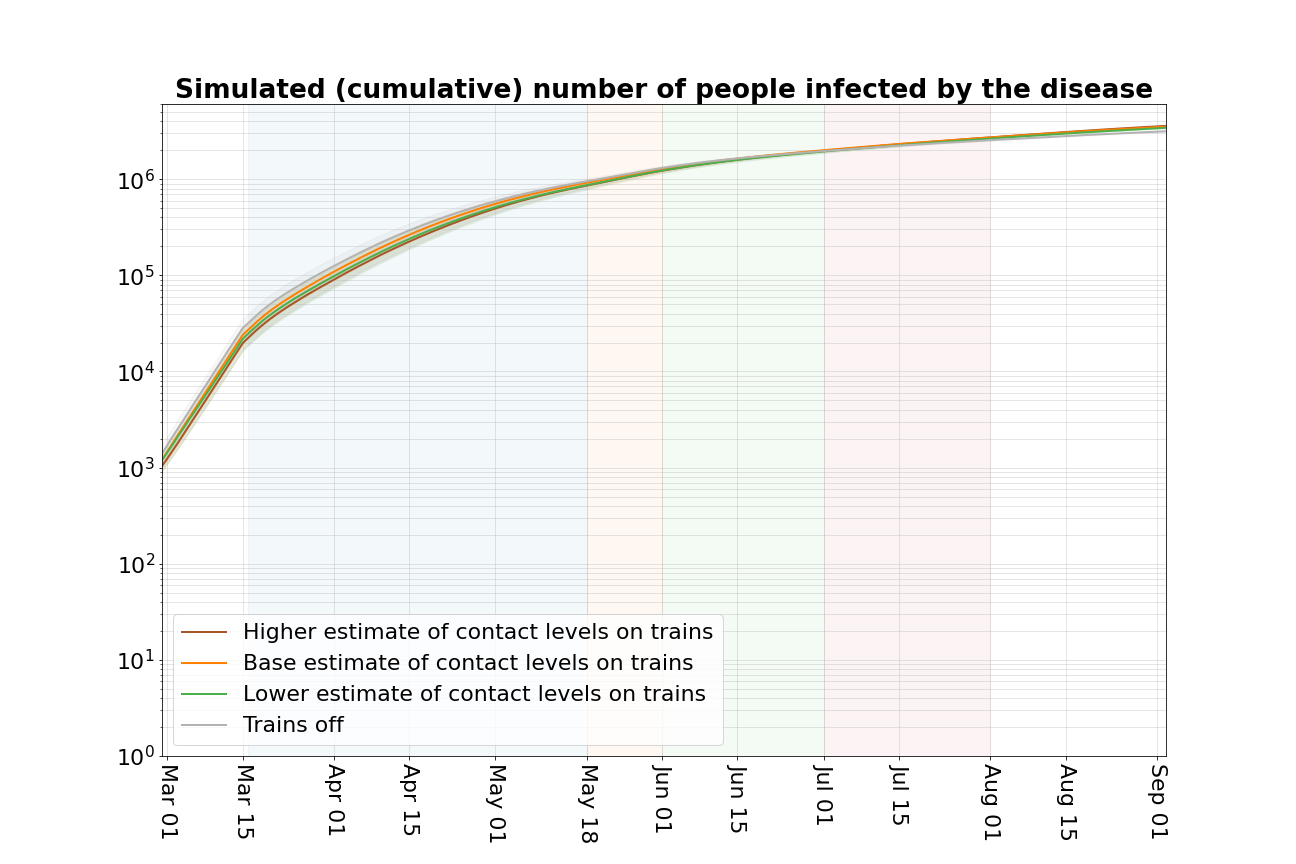}{Cumulative number of people infected by the disease on a linear and log scale.}{figure-affected-trains}

  \putfigures{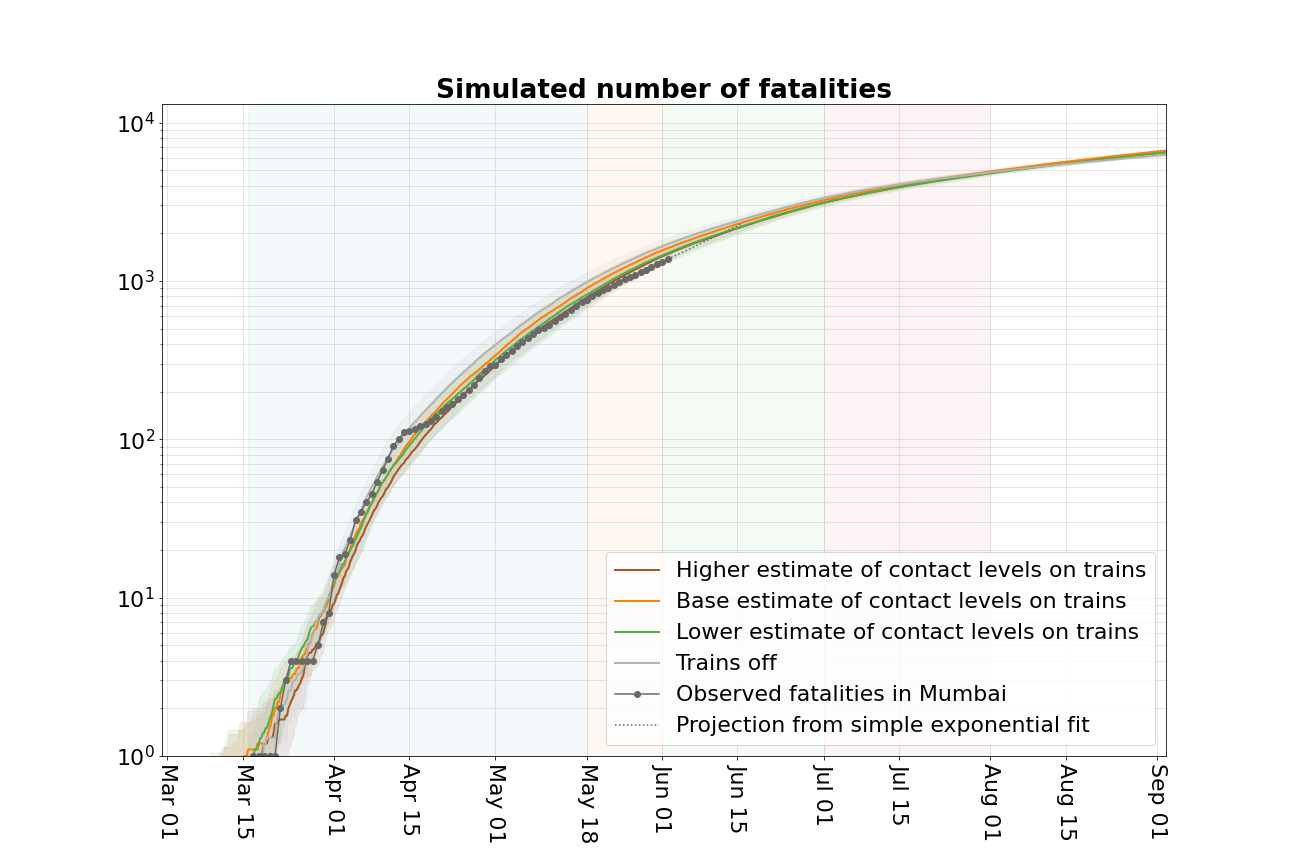}{Cumulative fatality numbers on a linear and log scale along with the observed number of fatalities in Mumbai.}{figure-fatal-trains}
  \begin{figure}
    \ContinuedFloat
    \centering
    \begin{subfigure}[h]{1.0\linewidth}
      \includegraphics[width=\linewidth]{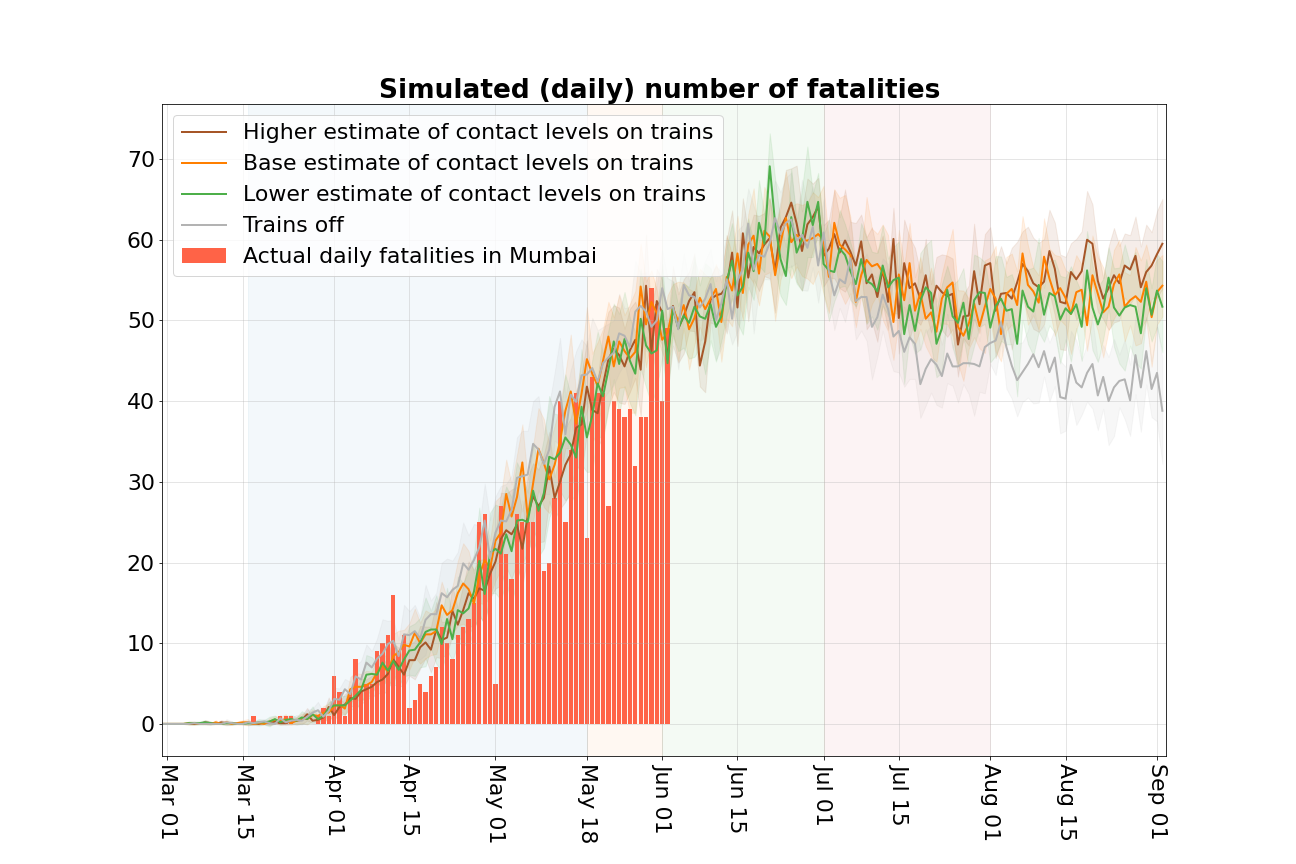}
      \caption{Simulated number of daily fatalities along with the actual
        daily statistics in Mumbai} \label{figure-dailyfatal-trains}
    \end{subfigure}
    \caption{\mycaption}
  \end{figure}
}


}

\end{enumerate}

\section*{Acknowledgments}

We sincerely thank Shubhada Agarwal, Siddharth Bhandari, Anirban Bhattacharjee, Anand Deo, and Poonam Kesarwani for their valuable comments and help.

\bibliographystyle{IEEEtran}
{
\bibliography{IEEEabrv,../covid19}
}

\end{document}